\documentclass{amsproc}
\usepackage{times,epsfig,amssymb}
\usepackage[mathcal]{euscript}

\DeclareSymbolFont{operators} {OT1}{ptmcm}{m}{n}
\DeclareSymbolFont{letters} {OML}{ptmcm}{m}{it}
\DeclareSymbolFont{symbols} {OMS}{pzccm}{m}{n}
\DeclareSymbolFont{largesymbols}{OMX}{psycm}{m}{n}

\DeclareMathAlphabet{\mathsf}{OT1}{phv}{m}{n}
\DeclareMathAlphabet{\mathrm}{OT1}{ptm}{m}{n}

\DeclareSymbolFont{ER}{U}{eur}{m}{n}
\DeclareSymbolFont{SY}{U}{psy}{m}{n}

\DeclareMathSymbol{,}{\mathpunct}{SY}{'054}
\DeclareMathSymbol{.}{\mathpunct}{SY}{'056}
\DeclareMathSymbol{:}{\mathpunct}{SY}{'072}
\DeclareMathSymbol{(}{\mathopen}{SY}{'050}
\DeclareMathSymbol{)}{\mathclose}{SY}{'051}
\DeclareMathSymbol{+}{\mathbin}{SY}{'053}
\DeclareMathSymbol{-}{\mathbin}{SY}{'055}
\DeclareMathSymbol{=}{\mathbin}{SY}{'075}
\DeclareMathSymbol{<}{\mathbin}{SY}{'074}
\DeclareMathSymbol{>}{\mathbin}{SY}{'076}
\DeclareMathSymbol{\leq}{\mathbin}{SY}{'243}
\DeclareMathSymbol{\geq}{\mathbin}{SY}{'263}
\DeclareMathSymbol{\nneq}{\mathbin}{SY}{'271}
\DeclareMathSymbol{\in}{\mathbin}{SY}{'316}
\DeclareMathSymbol{\nnotin}{\mathbin}{SY}{'317}
\DeclareMathSymbol{\times}{\mathbin}{SY}{'264}
\DeclareMathSymbol{\pm}{\mathbin}{SY}{'261}
\DeclareMathSymbol{\subset}{\mathbin}{SY}{'314}
\DeclareMathSymbol{\supset}{\mathbin}{SY}{'311}
\DeclareMathSymbol{\subseteq}{\mathbin}{SY}{'315}
\DeclareMathSymbol{\supseteq}{\mathbin}{SY}{'312}
\DeclareMathSymbol{/}{\mathord}{SY}{'057}
\DeclareMathSymbol{\ast}{\mathord}{SY}{'052}
\DeclareMathSymbol{\perp}{\mathord}{SY}{'136}

\renewcommand{\neq}{\nneq}

\DeclareMathSymbol{\emptyset}{\mathord}{SY}{'306}
\DeclareMathSymbol{\oplus}{\mathord}{SY}{'305}

\numberwithin{equation}{section}

\newcommand{\Z}{\mathbb{Z}}

\newcommand{\R}{\mathbb{R}}

\renewcommand{\S}{\mathbb{S}}

\newcommand{\cG}{\mathcal{G}}
\newcommand{\cH}{\mathcal{H}}

\newcommand{\cP}{\mathcal{P}}

\newcommand{\cT}{\mathcal{T}}

\newcommand{\tr}{\mathrm{tr}}

\DeclareMathOperator{\Prob}{\mathrm{Prob}}
\newcommand{\1}{\mathbb{I}}
\newcommand{\spec}{{\ensuremath{\rm spec}}}

\renewcommand{\det}{\mathrm{det\ }}

\newcommand{\vol}{\mathrm{vol}}
\newcommand{\enn}{\mathfrak{n}}
\newcommand{\ennn}{\mathsf{n}}

{\bf}{\it}
{\bf}{\it}
{\bf}{\it}
{\bf}{\it}
{\it}{\rm}
{\bf}{\it}
{\it}{\rm}
{\bf}{\it}

\begin{document}

\title[Statistical ensembles and density of states]
{Statistical ensembles and density of states}

\author{Vadim Kostrykin}
\address{Vadim Kostrykin, Fraunhofer-Institut f\"{u}r Lasertechnik,
Steinbachstra{\ss}e 15, D-52074 Aachen, Germany}
\email{kostrykin@t-online.de, kostrykin@ilt.fhg.de}

\author{Robert Schrader}
\address{Robert Schrader, Institut f\"{u}r Theoretische Physik,
Freie Universit\"{a}t Berlin, Arnimallee 14,
D-14195 Berlin, Germany} \email{schrader@physik.fu-berlin.de}
\thanks{The second author is supported in part by
DFG SFB 288 ``Differentialgeometrie und Quantenphysik''}

\subjclass{(2000 Revision) Primary 47A40, 82B10; Secondary 82B05, 82D20}

\date{February 18, 2002}

\dedicatory{Dedicated to Yakov Sinai and David Ruelle on the
occasion of their 65th birthday}

\keywords{Statistical mechanics, ensembles, scattering theory}

\begin{abstract}
We propose a definition of microcanonical and canonical statistical
ensembles based on the concept of density of states. This definition
applies both to the classical and the quantum case. For the
microcanonical case this allows for a definition of a
temperature and its fluctuation, which might be useful in the
theory of mesoscopic systems. In the quantum case the concept of
density of states applies to one-particle
Schr\"{o}dinger operators, in particular to operators with a periodic
potential or to random Anderson type models. In the case of periodic
potentials we show that for the resulting $n$-particle system the
density of states is $[(n-1)/2]$ times
differentiable, such that like for classical microcanonical ensembles
a (positive) temperature may be defined whenever $n\geq 5$.
We expect that a similar result should also hold for Anderson type
models. We also provide the first terms in asymptotic
expansions of thermodynamic quantities at large energies for the microcanonical
ensemble and at large temperatures for the canonical ensemble.
A comparison shows that then both formulations asymptotically give the
same results.
\end{abstract}

\maketitle

\section{Introduction}

The classical concepts of canonical and grand canonical statistical ensembles are
well known to have important quantum analogs (see, e.g., standard textbooks like
\cite{LL,KHuang,R}). To the best of our knowledge the notion of a microcanonical
ensemble in quantum theory has not received the attention we think is deserves. This
is unfortunate in view of recent developments in the classical and quantum theory of
small ensembles. So far the main motivation for a microcanonical analysis of small,
classical systems came from gravitational physics with its long range forces (see,
e.g., \cite{Antonov,Antonov1,LB,T,Hertel:Narnhofer:Thirring,LW,Padmanabhan} and the
references quoted there) and from nuclear physics with the associated fragmentation
processes (see, e.g., \cite{Gr,Gr1} and the references therein). The main aim was to
ask for the most probable distribution and to view this as an equilibrium phenomenon.
By energy conservation for closed systems it is natural to consider equilibria in the
microcanonical description, where the temperature $T$ and its fluctuation are derived
quantities. For the operational definition of a temperature for small systems one
cannot use a heat bath but rather a small thermometer, see \cite[p.351]{T}.

Usually one introduces the notion of statistical ensembles to describe the systems
with large number of particles. Our approach will in some sense go in a different
direction. We deal with the question how small the system can be to allow a
thermodynamic description.

We recall that in the familiar classical microcanonical description $1/T$ may be
written as a mean over the energy shell of a certain function in the (finite
dimensional) phase space (see relation \eqref{tempav} below). Considering the square
of this function one may in addition consider fluctuations of the inverse
temperature. In case the ergodic hypothesis holds such averages may then be written
as time averages using the dynamics of the classical Hamilton function (see, e.g.,
\cite{Rugh1,Rugh2,Ba,NS}). An important feature of a microcanonical ensemble is that
the temperature may decrease with energy giving rise to a negative heat capacity. The
first references which envisage such a situation seem to be \cite{LL} and \cite{tH}.
The example considered in \cite{LL} is that of a star, which has used up its nuclear
fuel and then heats up under gravitational contraction. A first model discussion for
a supernovae providing a negative specific heat was given in \cite{T}. For recent
experiments on small systems, see, e.g., \cite{Gr,Sch,Sch1,Italiener} and references
given there. For more recent theoretical discussions of small systems, see, e.g.,
\cite{IC,MPT}. We mention also the recent work \cite{MPT} where a system of classical
particles with power law potentials was considered. The results of this work indicate
that already for particle numbers larger than $10$ there might be a relaxation to the
microcanonical equilibrium starting from an arbitrary initial state.

It is the purpose of this article to provide a useful quantum theoretical approach to
the theory of canonical and microcanonical ensembles which may be applied to small
(i.e., mesoscopic) systems. This will include a discussion of fluctuations, the heat
capacity and their interrelation. We also compare the microcanonical and the
canonical ensemble. Our approach will be based on the theory of (integrated) density
of states, which in the quantum case is a well known notion in solid state physics
(see, e.g., \cite{AM,M}). This theory has been investigated intensively in the last
years by mathematical physicists in the context of one-particle Schr\"{o}dinger
operators, in particular, operators with a periodic potential and random Schr\"{o}dinger
operators like Anderson type models. For recent references consult, e.g.,
\cite{KS1,CHN,KS2,Hislop:Klopp} and for references before 1992
\cite{GLP,WKirsch,CL,Pastur:Figotin}.

In our approach the microcanonical description is given by defining the entropy to be
$k$ (the Boltzmann constant) times the logarithm of the density of states, so that
the entropy comes to be a function of the energy. As usual the inverse temperature
$1/T$ is then the derivative of the entropy, i.e., $k$ times the logarithmic
derivative of the density of states. The canonical description on the other hand is
by definition given in terms of a partition function, now defined to be the
Stieltjes-Laplace transform of the integrated density of states w.r.t.\ the variable
$\beta=1/kT$ canonically conjugate to the energy.

In particular we will address the familiar question under which conditions these two
descriptions give approximately the same answer. In fact, we will see that the well
known methods of comparison in the standard formulation easily carry over to this new
formulation. In addition we will show, both in the classical and quantum cases, that
for high energies or correspondingly high temperatures the microcanonical and the
canonical description give the same results.

So for our approach to work in the microcanonical case, differentiability of the
integrated density of states up to third order is necessary. Now our main observation
is that smoothness of the integrated density of states for the resulting $n$-particle
theory increases with $n$. This will in particular allow us to consider
microcanonical ensembles for particles moving in a periodic external potential
provided $n\ge 5$. The reason is that for one-particle Schr\"{o}dinger operators with a
periodic potential the density of states (which is the derivative of the integrated
density of states) is well known to exhibit so called van Hove singularities in space
dimensions greater or equal to two (see, e.g., \cite{AM,M}). In one dimension the
integrated density of states itself has square root singularities. These are smoothed
out when going to higher particle number. We expect a similar property to hold for
particles moving in an external random potential. We will argue why increased
smoothness with $n$ is related to the well known increasing smoothness of the
integrated density of states with the dimension of the space. For Anderson-type
random Schr\"{o}dinger operators the existence of the density of states (i.e., the
absolute continuity of the integrated density of states) is a complicated, still not
completely solved problem. Very little known is known about regularity properties of
the density of states \cite{CH,CHN,KS2,Hislop:Klopp}.

In addition we will invoke notions from scattering theory and, in particular, the
scattering phase shift (or total scattering phase) at a fixed energy $E$. It is an
old observation of Beth and Uhlenbeck \cite{BU} that the second virial coefficient in
statistical mechanics is related to the phase shift (see also \cite{KHuang} and
relations \eqref{Krein} and \eqref{Krein1} below). We will use the concept of the
phase shift density introduced in \cite{KS0,KS1} to give the asymptotic behavior at
large energies (or temperatures) of the main thermodynamic quantities. In particular,
in \cite{KS0,KS1} we established that up to a factor of $\pi$ this phase shift
density is the difference of the integrated density of states with potential and the
free theory. This relation is analogous to the change of the number of particle
states found by Friedel \cite{Fr1,Fr2} in the case of a single impurity. The phase
shift density will be used to establish relations on the shift of the temperature
induced by the potential in the microcanonical context and on the shift of the mean
energy in the canonical context. Furthermore, we establish a relation on the shift of
the mean energy density for a system of noninteracting electrons moving in a periodic
or a random potential. This relation is analogous to a theorem of Fumi \cite{Fu}
which relates the shift of the ground state energy due to a single impurity to the
scattering phase.

We will not discuss situations where in addition to energy other quantities, like,
e.g., angular momentum, are conserved. Also we will not cover the situation where in
addition to an external potential there is interaction between the particles. This is
of course an important issue worth pursuing. We note that non-relativistic
$n$-particle scattering theory with all possible fragmentation and bound state
channels has been extensively analyzed (see, e.g., \cite{GS,DG,HS} and the references
quoted there). We expect that some of our results may be extended to this situation,
although at present it is unclear to us, which r\^{o}le such multichannel spectral
properties play in the theory of integrated density of states.

In addition, it would be interesting to see whether the present approach to the
theory of quantum mechanical, microcanonical ensembles could serve as a laboratory
for a fresh look at ergodic theory in quantum mechanics (see, e.g., \cite{N1,FP,F}).

The article is organized as follows. In Section \ref{sec:2} we will review the
classical theory of microcanonical and canonical ensembles in a form which serves as
a motivation and comparison for our quantum mechanical approach to be given in
Section \ref{sec:3}. For the classical models we consider, we will introduce a
classical notion of (integrated) density of states and formulate the resulting theory
of statistical ensembles. We will see that in this classical setup differentiability
indeed increases with the particle number. In addition, we will provide a new example
for a system with negative heat capacity. Also we will provide an example where the
microcanonical and the canonical descriptions asymptotically for large $E$ and $T$
give the same (mean) energy -- temperature relation. Recall that in the usual
canonical description the heat capacity is always positive.

So the discussion in Section \ref{sec:2} will be rather extensive, the reason being
that the quantum theory is then simply obtained by replacing the classical
(integrated) density of states by the corresponding quantum version.

Section \ref{sec:3} will contain the corresponding quantum mechanical formulation of
both the microcanonical and the canonical ensemble using the quantum notion of the
integrated density of states. For large energies we again will argue on the basis of
a yet unproven conjecture on the high energy behavior of the phase shift \cite{S}
that the (mean) energy -- temperature relations are asymptotically equal for both
ensembles. As already mentioned we view this result as an indication of the
reasonableness of our thermodynamic approach to small systems. Section \ref{sec:3}
will also include a brief comparison between the classical and the quantum theory for
small $\hbar$. Finally in Section \ref{sec:3} we also briefly discuss the grand
canonical ensemble in terms of the integrated density of states.

In the appendix we will show in the quantum context for periodic potentials that
singularities of the $n$-particle density of states are smoothed out with increasing
number of particles $n$. In addition we will show in the classical case how the
randomness in stochastic potentials smoothes out the integrated density of states.

We avoid to formulate our results as theorems and propositions. However making any
statement we provide a discussion of what is rigorously proven and what is only
conjectured motivated by the physical intuition.

\textbf{Acknowledgments.} We are indebted to R.M.~Dudley, P.D.~Hislop,
M.~Karowski, H.~Narnhofer, M.~Schmidt, K.-D.~Schotte, H.~Spohn, and W.~Thirring for
valuable comments. We thank D.H.E.~Gross and O.~Fliegans for providing references.

\section{The classical theory}\label{sec:2}
\setcounter{equation}{0}

In this section and for the purpose of comparison and as part of our motivation we
will briefly review the well known concepts of microcanonical and canonical ensembles
in the classical case. Some of the material, however, seems to be new. In particular
we will provide new examples with negative heat capacity. Previous examples in the
context of gravitation theory are given, e.g., \cite{LW,T} and the references quoted
there.

In classical theory the starting point is a finite dimensional phase space $\cP$
which is a symplectic manifold. We will assume, in addition, that $\cP$ also is a
K\"{a}hler manifold, i.e., it is also a Riemannian manifold and the symplectic and
Riemanian structures are compatible. This implies, in particular, that the Liouville
volume form and Riemannian volume form agree. We denote this volume form by
$d\vol_{\cP}$. Also a Hamilton function $H_{\mathrm{cl}}$ on $\cP$ is supposed to be
smooth and bounded below in case $\cP$ is not compact.

First we assume $\cP_{\leq E}=\{p\in\cP|\; H_{\mathrm{cl}}(p)\leq E\}$ is compact for
all $E$. Note that Hamilton functions of the form $H_{\mathrm{cl}}(p,x)=p^2/2m +
V(x)$ with bounded $V$ do not satisfy this assumption. We will turn to this later.

The function
\begin{equation}\label{nc1}
N(E;H_{\mathrm{cl}})=\int_{\cP_{\le E}}d\vol_{\cP}
  =\int_{\cP}\Theta(E-H_{\mathrm{cl}})d\vol_{\cP},
\end{equation}
where $\Theta$ is the Heaviside step function, is increasing w.r.t.\ the energy $E$.
Its derivative is given as
\begin{equation*}
0\le W(E;H_{\mathrm{cl}})=\frac{d}{dE}N(E;H_{\mathrm{cl}})
  =\int_{\cP}\delta(E-H_{\mathrm{cl}})d\vol_{\cP}
\end{equation*}
with $\delta$ being the Dirac $\delta$-function. The associated microcanonical
ensemble is then given by defining the entropy as a function of the energy $E$ as
$S(E;H_{\mathrm{cl}})=k\ln W(E;H_{\mathrm{cl}})$ and the temperature
$T(E)=T(E;H_{\mathrm{cl}})$, again considered as a function of the energy $E$, is
then defined by
\begin{equation}\label{t1}
\frac{1}{T(E)}=\frac{d S(E;H_{\mathrm{cl}})}{dE}=
k\frac{d}{dE}\ln\frac{d N(E;H_{\mathrm{cl}})}{dE}
=k\frac{1}{W(E;H_{\mathrm{cl}})}\frac{dW(E;H_{\mathrm{cl}})}{dE}
\end{equation}
with $k$ being the Boltzmann constant. So provided the r.h.s.\ of
\eqref{t1} is meaningful the temperature is defined but possibly negative. In standard
situations $W(E;H_{\mathrm{cl}})$ increases with $E$ such that $T$ is nonnegative.
However, there are also situations, where the temperature $T$ may become negative
(see, e.g., \cite{LL}).

The function $W(E;H_{\mathrm{cl}})$ and its two first derivatives may be written in
the form
\begin{eqnarray}\label{rie}
W(E;H_{\mathrm{cl}})&=&\int_{\cP_E}d\mu_E\nonumber\\
\frac{d}{dE}W(E;H_{\mathrm{cl}})&=
&\int_{\cP_E}\nabla\cdot
\frac{\nabla H_{\mathrm{cl}}}{|\nabla H_{\mathrm{cl}}|^2}d\mu_E\\
\frac{d^2}{dE^2}W(E;H_{\mathrm{cl}})&=
&\int_{\cP_E}\nabla\cdot
\left(\left(\nabla\cdot\frac{\nabla H_{\mathrm{cl}}}{|\nabla H_{\mathrm{cl}}|^2}\right)
\frac{\nabla H_{\mathrm{cl}}}{|\nabla H_{\mathrm{cl}}|^2}\right)d\mu_E\nonumber
\end{eqnarray}
with the following notation. The manifold $\cP_E$, the boundary of $\cP_{\le E}$, is
the energy surface for the energy $E$, i.e., the set of points $p\in \cP$ for which
$H_{\mathrm{cl}}(p)=E$. The measure $d\mu_E$ on $\cP_E$ is given as $d\mu_E=|\nabla
H_{\mathrm{cl}}|^{-1}d\vol_{\cP_E}$, where $d\vol_{\cP_E}$ is the canonical volume
form on $\cP_E$ given by the Riemannian metric. $\nabla$ is the covariant gradient
and $|\nabla H_{\mathrm{cl}}|(p)$ the length of the vector $\nabla
H_{\mathrm{cl}}(p)$, which is normal to the surface $\cP_E$ at $p\in\cP_E$. So
provided $\nabla H_{\mathrm{cl}}(p)\neq 0$, the vector $1/|\nabla
H_{\mathrm{cl}}(p)|\;\nabla H_{\mathrm{cl}}(p)$ is the outward unit normal vector to
$\cP_E$ at $p\in\cP_E$, i.e., this vector points into the complement $\cP_{>E}$ of
$\cP_{\le E}$. Near $\cP_{E}$ one has the familiar relation $d\vol_{\cP}=d\mu_E dE$,
whence the first relation in \eqref{rie}. Finally $\nabla\cdot$ is the covariant
divergence. The second and third relation in
\eqref{rie} are obtained from the first by (repeated) use of the theorem of Gauss
(see \eqref{gauss} below). Higher derivatives may be calculated similarly. In
standard situations $\cP_E$ is a compact set, generically of codimension 1. For
$W_{\mathrm{cl}}(E;H_{\mathrm{cl}})$ and its derivatives to be well defined one has
to assume that $\nabla H_{\mathrm{cl}}(p)\neq 0$ for $p\in \cP_E$ (or at least that
suitable inverse powers of $|\nabla H_{\mathrm{cl}}|(p)$ are integrable over
$\cP_E$). Then in particular $\cP_E$ is a smooth submanifold of $\cP$ of codimension
1 and $W(E;H_{\mathrm{cl}})\neq 0$ provided the energy $E$ is such that $\cP_E\neq
\emptyset$. The first relation in \eqref{rie} dates back to Khinchin \cite{Khinchin}.

The temperature $T(E;H_{\mathrm{cl}})$ defined by \eqref{t1} is an increasing
function at $E$ if
\begin{equation}
\label{ineq1}
 W(E;H_{\mathrm{cl}})\frac{d^2}{dE^2}W(E;H_{\mathrm{cl}})\leq
\left(\frac{d}{dE}W(E;H_{\mathrm{cl}})\right)^2.
\end{equation}
We may rephrase this as follows (see also \cite{Rugh1,Rugh2}). Let
\begin{equation*}
\langle f\rangle_E= \frac{\int_{\cP_E} fd\mu_E}
{\int_{\cP_E} d\mu_E}=\frac{\int
  \delta(E-H_{\mathrm{cl}})fd\vol_{\cP}}{\int
  \delta(E-H_{\mathrm{cl}})d\vol_{\cP}}
\end{equation*}
denote the average over the energy shell $\cP_E$ of a real valued function $f$
defined on $\cP_E$. For given $H_{\mathrm{cl}}$ consider in particular the function
$\cT$ on $\cP$ of the dimension of energy given as
\begin{equation*}
{\cT}^{-1}=\nabla\cdot\frac{\nabla H_{\mathrm{cl}}}{|\nabla H_{\mathrm{cl}}|^2}
\end{equation*}
such that
\begin{equation}
\label{tempav}
\frac{1}{T(E)}=k\langle {\cT}^{-1}\rangle_E.
\end{equation}
In general $kT(E)\neq \langle \cT\rangle_{E}$. Consider for example the case where
$0<\cT<\infty$ on $\cP_E$. Then
\begin{equation*}
1=\langle 1\rangle_E=\langle\cT^{1/2}\cT^{-1/2}\rangle_E
\leq\langle\cT\rangle_E^{1/2}\langle\cT^{-1}\rangle_E^{1/2}
\end{equation*}
by the Schwarz inequality giving $\langle
\cT\rangle_E\ge kT(E)$ with equality if and only if $\cT$ is constant on $\cP_E$.

We write
\begin{equation}\label{G}
\begin{split}
& \nabla\cdot\left(\left(\nabla\cdot\frac{\nabla H_{\mathrm{cl}}}{|\nabla
      H_{\mathrm{cl}}|^2}\right)
\frac{\nabla H_{\mathrm{cl}}}{|\nabla H_{\mathrm{cl}}|^2}\right)
=\frac{1}{\cT^2}+\cG,\\
& \cG=\left(\nabla\left(\nabla\cdot\frac{\nabla H_{\mathrm{cl}}}{|\nabla
      H_{\mathrm{cl}}|^2}\right)\right)\cdot
\frac{\nabla H_{\mathrm{cl}}}{|\nabla H_{\mathrm{cl}}|^2}
=(\nabla{\cT}^{-1})\cdot\frac{\nabla H_{\mathrm{cl}}}{|\nabla
      H_{\mathrm{cl}}|^2}.
\end{split}
\end{equation}
Here $\:\cdot\:$ denotes the scalar product of two vectors (or rather vector fields)
given by the Riemannian metric, such that in particular $|\nabla
H_{\mathrm{cl}}|^2=\nabla H_{\mathrm{cl}}\cdot\nabla H_{\mathrm{cl}}$.

Consider the quantity $\Delta(T^{-1})(E)=\Delta(T^{-1})(E;H_{\mathrm{cl}})$ given by
\begin{equation}
\label{fluc}
\Delta(T^{-1})(E)^2=k^2\left(\langle {\cT}^{-2}\rangle_{E}-
\langle {\cT}^{-1}\rangle_{E}^2\right)\ge 0
\end{equation}
may be viewed as the fluctuation of the inverse temperature at energy $E$. Note that
the inequality is a consequence of the Schwarz inequality or equivalently of the
familiar relation $0\le\langle(f-\langle f\rangle_E)^2\rangle_E$. In particular
$\Delta(T^{-1})(E)$ vanishes if and only if $\cT$ is constant on $\cP_E$. The
fluctuation and the heat capacity $c_{\mathrm{v}}(E)\equiv
c_{\mathrm{v}}(E;H_{\mathrm{cl}})$ defined by
\begin{equation}
\label{heatcap}
\frac{1}{c_{\mathrm{v}}(E)}=\frac{d}{dE}T(E)=-T(E)^2\frac{d}{dE}\frac{1}{T(E)}=
-kT(E)^2\frac{d}{dE}\ln\frac{d}{dE}W(E;H_{\mathrm{cl}})
\end{equation}
are related by
\begin{equation}
\label{heatvarcap}
\Delta(T^{-1})(E)^2+\frac{k}{T(E)^2c_{\mathrm{v}}(E)}=-k^2\langle\cG\rangle_E.
 \end{equation}
Therefore condition \eqref{ineq1}, which guarantees that $c_{\mathrm{v}}(E)\ge 0$,
may be recast into the equivalent form
\begin{equation}
\label{ineq0}
\frac{1}{k^2}\Delta(T^{-1})(E)^2\le -\langle \cG\rangle_{E}.
\end{equation}
By \eqref{G} we have $|\nabla H_{\mathrm{cl}}|\cG=\nabla(1/\cT)\cdot|\nabla
H_{\mathrm{cl}}|^{-1}\nabla H_{\mathrm{cl}}$. This has the following geometric
interpretation. $|\nabla H_{\mathrm{cl}}|\cG$ is the component of the gradient of
$1/\cT$ in the normal direction. So if this gradient always points into $\cP_{\le E}$
(i.e., $\nabla (1/\cT)\cdot\nabla H_{\mathrm{cl}}(s)\le 0$ for all $s\in\cP_{E}$)
thus making $\cG$ negative (or zero) there, then the inequality \eqref{ineq0} is
satisfied for the energy $E$. Since $\nabla(1/\cT)=-(1/\cT)^2\nabla\,\cT$ this is
equivalent to the condition that the gradient of $\cT$ always points into
$\cP_{>E}=\cP\setminus \cP_{\leq E}$. Conversely, if the inequality in \eqref{ineq0}
is reversed and in particular if $\langle \cG\rangle_{E}
> 0$ (meaning that in the mean the gradient of $1/\cT$ points into $\cP_{>E}$, then
$T(E,H_{\mathrm{cl}})$ is a monotone decreasing function in $E$, i.e., the heat
capacity is negative.

By means of the Gauss theorem and using \eqref{G} the relations
\eqref{rie} can be written equivalently as follows
\begin{eqnarray}\label{gauss}
W(E;H_{\mathrm{cl}})&=&\int_{\cP_{\le E}}{\cT}^{-1}d\vol_{\cP},\nonumber\\
\frac{d}{dE}W(E;H_{\mathrm{cl}})&=
&\int_{\cP_{\le
    E}}\left({\cT}^{-1}+\cT\cG\right){\cT}^{-1}d\vol_{\cP},\\
\frac{d^2}{dE^2}W(E;H_{\mathrm{cl}})&=
&\int_{\cP_{\le E}}\left({\cT}^{-2}+2\cG +\cT\nabla\cG\cdot\frac{\nabla
H_{\mathrm{cl}}}{|\nabla H_{\mathrm{cl}}|^2}\right) {\cT}^{-1}d\vol_{\cP}.\nonumber
\end{eqnarray}
Moreover, for any differentiable function $f$ on $\cP_{\le E}$ the average over the
energy shell can be written as
\begin{equation}\label{gauss1}
\langle\,f\,\rangle_E=\frac{\int_{\cP_{\le E}}
\left(\nabla\,f\cdot\frac{\nabla\,H_{\mathrm{cl}}}{|\nabla\,H_{\mathrm{cl}}|^2}
+f{\cT}^{-1}\right)d\vol_{\cP}} {\int_{\cP_{\le E}}{\cT}^{-1}d\vol_{\cP}}.
\end{equation}

Assume now that $\cT >0$ on $\cP_{\le E}$. Then we may introduce the probability
measure on $\cP_{\le E}$ by
\begin{equation}\label{dnu}
d\nu_{E}=\frac{1}{W(E;H_{\mathrm{cl}})}\Theta(E-H_{\mathrm{cl}}){\cT}^{-1}d\vol_{\cP}.
\end{equation}
In analogy to $\langle\cdot\rangle_E$ let $\langle\cdot\rangle_E^{\prime}$ denote the
resulting mean. Then relations \eqref{tempav} can be rewritten as
\begin{equation*}
\frac{1}{T(E;H_{\mathrm{cl}})}=k\langle {\cT}^{-1}+\cT\cG\rangle_E^{\prime}.
\end{equation*}
However, the resulting inverse temperature fluctuation given by
\begin{equation}\label{flucprime}
\Delta^{\prime}(T^{-1})(E)
=\left(k^2\langle({\cT}^{-1}+\cT\cG)^2\rangle_{E}^{\prime}
-T(E)^{-2}\right)^{1/2}
\end{equation}
is in general different from $\Delta(T^{-1})(E)$. In fact, by \eqref{gauss1} and the
Gauss theorem
\begin{equation*}
0\le\langle{\cT}^{-2}\rangle_E=
\langle({\cT}^{-2}+2\cG)\rangle_E^{\prime}
=\langle({\cT}^{-1}+\cT\cG)^2\rangle_E^{\prime}-
\langle (\cT\cG)^2\rangle_E^{\prime}.
\end{equation*}
This implies $\Delta(T^{-1})(E)\le \Delta^{\prime}(T^{-1})(E)$ which is an equality
if and only if $\cT\cG$ vanishes identically on $\cP_{\le E}$. Relation
\eqref{heatvarcap} is now replaced by
\begin{equation*}
\Delta^{\prime}(T^{-1})(E)^2+\frac{k}{T(E)^2c_{\mathrm{v}}(E)}=-\langle
\cT\nabla\cG\cdot\frac{\nabla H_{\mathrm{cl}}}{|\nabla
  H_{\mathrm{cl}}|^2}
\rangle^{\prime}_{E},
\end{equation*}
so alternatively to \eqref{ineq0} the condition
\eqref{ineq1} for the heat capacity to be positive can now be written as
\begin{equation*}
\frac{1}{k^2}\Delta^{\prime}(T^{-1})(E)^2\le
-\langle
\cT\nabla\cG\cdot\frac{\nabla H_{\mathrm{cl}}}{|\nabla H_{\mathrm{cl}}|^2}\rangle^{\prime}_{E}.
\end{equation*}

The alternative definition \eqref{flucprime} of the inverse temperature fluctuation
has the disadvantage that besides the assumed positivity of $\cT$ it involves the
values of $H_{\mathrm{cl}}$ on all of $\cP_{\le E}$. In contrast and due to the
appearance of derivatives of $H_{\mathrm{cl}}$, the definition \eqref{fluc} only
involves the values of $H_{\mathrm{cl}}$ near $\cP_E$.

The Boltzmann-Gibbs partition function in the associated canonical ensemble may also
be given in terms of the functions $N(E;H_{\mathrm{cl}})$ and $W(E;H_{\mathrm{cl}})$
as
\begin{equation}
\label{ccan}
Z(\beta;H_{\mathrm{cl}})=\int e^{-\beta H_{\mathrm{cl}}}d\vol_{\cP}
     =\int e^{-\beta E}dN(E;H_{\mathrm{cl}})
     =\int e^{-\beta E}W(E;H_{\mathrm{cl}})dE, \quad\beta=1/kT.
\end{equation}
The (mean) energy-temperature relation now takes the form
\begin{equation*}
\bar{E}(\beta;H_{\mathrm{cl}})=-\frac{d}{d\beta}Z(\beta;H_{\mathrm{cl}})
               =\langle E\rangle_{\beta}\quad\text{with}\quad
               \langle g\rangle_{\beta}=
\frac{\int g(E)e^{-\beta E}W(E;H_{\mathrm{cl}})dE}{Z(\beta;H_{\mathrm{cl}})}
\end{equation*}
being the mean of any function $g$ of the energy for given $\beta$. The mean energy
now is a monotone increasing function of temperature. Since $\langle
E^2\rangle_{\beta}\geq\langle E\rangle^2_{\beta}$ by the Schwarz inequality,
$\Delta\bar{E}(\beta;H_{\mathrm{cl}})=(\langle E^2\rangle_{\beta}-\langle
E\rangle^2_{\beta})^{1/2}$ gives the energy fluctuation. Stated differently, the heat
capacity defined in this canonical context as
\begin{equation}
\label{canheatcap}
c_{V}(\beta;H_{\mathrm{cl}})=
\frac{\partial}{\partial
  T}\bar{E}(\beta;H_{\mathrm{cl}})
=-\frac{\beta^2}{k}\frac{\partial}{\partial \beta}
\bar{E}(\beta;H_{\mathrm{cl}})
\end{equation}
is always nonnegative.

The following discussion (see also \cite{Rugh2}) relates the microcanonical and
canonical descriptions. It is an adaptation of well known arguments employed in this
context (see, e.g., \cite{KHuang}) and is recalled for later purpose. Write
\begin{equation}
\label{ccan1}
Z(\beta;H_{\mathrm{cl}})
     =\int e^{-\beta E+\ln\frac{1}{\beta}W(E;H_{\mathrm{cl}})}dE.
\end{equation}
The integrand takes an extremal value at $E_{\max}=E_{\max}(T)$ given implicitly by
\begin{equation}
\label{ccan2}
\frac{1}{T}=\left.k\frac{d}{dE}\ln W(E;H_{\mathrm{cl}})\right|_{E=E_{\max}},
\end{equation}
which agrees with \eqref{t1}. If a solution $E_{\max}$ to the relation \eqref{ccan2}
exists it is unique. If the inequality \eqref{ineq1} holds for all $E$ and is a
strict inequality at $E=E_{\max}$ then the extremal value of the integrand is a
maximal one. Indeed, for $E$ near $E_{\max}$ we obtain
\begin{equation*}
-\beta E +\ln\left(\frac{1}{\beta}
 W(E;H_{\mathrm{cl}})\right)=-\frac{1}{2}\alpha(E-E_{\max})^2
+O((E-E_{\max})^3)
\end{equation*}
with $\alpha=\alpha(\beta;H_{\mathrm{cl}}^{(n)})$ given by $\alpha=-d^2\ln
W(E=E_{\max};H_{\mathrm{cl}})/dE^2$, i.e.,
\begin{equation}
\label{ccan4}
\alpha=\left.\frac{\left(\frac{d}{dE}W(E;H_{\mathrm{cl}})\right)^2
-\frac{d^2}{dE^2}W(E;H_{\mathrm{cl}})\cdot W(E;H_{\mathrm{cl}})}
{W(E;H_{\mathrm{cl}})^2}\right |_{E=E_{\max}}.
\end{equation}
which is nonnegative if \eqref{ineq1} holds and is strictly positive if \eqref{ineq1}
is a strict inequality at $E=E_{\max}$. In fact, in the microcanonical notation
$\alpha$ is related to the heat capacity via $\alpha=1/(k
c_{\mathrm{v}}(E_{\max})T(E_{\max})^2)$ by
\eqref{ccan2}. So when $\alpha >0$ is large, to a good approximation
the integrand
in \eqref{ccan1} is given by a Gaussian distribution with variance $\alpha^{-1}$. The
energy fluctuation is then $\Delta\bar{E}(\beta;H_{\mathrm{cl}})\sim\alpha^{-1/2}$.
Finally if in the canonical ensemble the heat capacity is defined as
$c_{\mathrm{v}}(T)=d\bar{E}(\beta;H_{\mathrm{cl}})/dT$ this quantity then agrees with
the microcanonical heat capacity at the energy $E_{\max}$, i.e.,
$kT^2c_{\mathrm{v}}(T)=\Delta\bar{E}(\beta;H_{\mathrm{cl}})^2\sim
kT^2c_{\mathrm{v}}(E_{\max})$.

>From now on we will consider classical Hamilton functions with non-compact $\cP_{\leq
E}$ such that the above discussion does nor apply. The following discussion serves as
a preparation to what we will do in the quantum case in the next section. More
concretely, we consider a particle of mass $m>0$ moving in $\R^d$ under the influence
of a periodic bounded potential $V(x)$, i.e., $V(x)=V(x+j)$ for all $x\in\R^d$ and
all $j\in \Z^d$. Thus the classical Hamilton function is given as
$H_{\mathrm{cl}}(x,p)=p^2/2m + V(x)$ and the phase space $\cP$ is $\R^{2d}$ with the
canonical symplectic structure. We will consider the resulting $n$-particle theory
with phase space $\R^{2nd}$ and classical Hamilton function
$H_{\mathrm{cl}}^{(n)}(x,p)=p^2/2m+V^{(n)}(x)=
H_{0\,\mathrm{cl}}^{(n)}(p)+V^{(n)}(x)$. Here
$p=(p_1,p_2,...,p_n),x=(x_1,x_2,...,x_n),\;p_i,x_i\in \R^d$ and
$V^{(n)}(x)=\sum_iV(x_i)$. For the purpose of this article it will suffice to assume
the potential to be bounded.

We will also consider random, classical Hamiltonians of the so called Anderson type.
More precisely, we assume the one particle random potential $V=V_{\omega}$ to be of
the form
\begin{equation}\label{stopo}
V_\omega(x)=\sum_{j\in \Z^d}q_{j}(\omega)V_0(x-j).
\end{equation}
Here $\omega=\{\omega_j\}_{j\in \Z^d}$ is an element of the probability space
$\Omega=\R^{\Z^d}$ with a probability measure $\Prob$ and
$q_{j}(\omega)=q(\omega_{j})$ for some real valued bounded measurable function $q$.
Therefore $\omega_j$ are independent random variables with the same distribution. In
other words there is a probability measure $d\mu$ on $\R$ such that $\Prob(q_{j}\in
I)=\int_{I}d\mu(q)$ for any interval $I\subset\R$ and any $j\in\Z^d$. Also we assume
that $V_0$ is a bounded function with support in the unit cube in $\R^d$, such that
$V_{0}(x-j)$ and $V_{0}(x-k)$ have non-overlapping support when $j\neq k$. This
results in a Hamiltonian $H_{\mathrm{cl}}(p,x)=H_{\mathrm{cl},\omega}(p,x)
=p^2/2m+V^{(n)}_{\omega}(x)$, again with the obvious notation
$V^{(n)}_{\omega}(x)=\sum_{i}V_{\omega}(x_{i})$.

In the considered cases the function $N(E;H_{\mathrm{cl}}^{(n)})$ defined by
\eqref{nc1} is infinite but the integrated density of states, given as
\begin{equation*}
\enn(E;H_{\mathrm{cl}}^{(n)})=\lim_{\Lambda\rightarrow\infty}\frac{1}{|\Lambda|^n}
\int_{\Lambda^n\otimes \R^{nd}}\Theta(E-H_{\mathrm{cl}}^{(n)}(p,x))\
d^{nd}x\ d^{nd}p
\end{equation*}
exists where $\Lambda\subset \R^{d}$ and
$\Lambda^n=\Lambda\times\ldots\times\Lambda\subset
\R^{nd}$. If $V$ is periodic we obviously have
\begin{equation}\label{cldensn}
\begin{split}
\enn(E;H_{\mathrm{cl}}^{(n)}) &=
\int_{\Lambda_0^n\otimes \R^{nd}}\Theta(E-H_{\mathrm{cl}}^{(n)}(p,x))\
d^{nd}x\ d^{nd}p\\ &=
\frac{1}{nd}\mathrm{Vol}(\S^{nd-1})
(2m)^{nd/2}\int_{\Lambda_0^n}
\Theta(E-V^{(n)}(x))(E-V^{(n)}(x))^{nd/2}d^{nd}x.
\end{split}
\end{equation}
where $\Lambda_0$ is the unit cube in $\R^d$, $\S^{nd-1}$ is the unit sphere in
$\R^{nd}$ and $\mathrm{Vol}(\S^{nd-1})$ its area. In the case of free motion, i.e.,
$V=0$, and using the fact $\mathrm{Vol}(\S^{nd-1})=2\pi^{nd/2}/\Gamma(nd/2)$ we
obtain
\begin{equation}\label{cldens0}
\enn(E;H_{0\,\mathrm{cl}}^{(n)})=\frac{1}{nd}\mathrm{Vol}(\S^{nd-1})
(2mE)^{nd/2}
=\frac{(2\pi m E)^{nd/2}}{\Gamma(nd/2+1)},\quad E\geq 0.
\end{equation}
Note that from \eqref{cldensn} it follows that the order of differentiability of
$\enn(E;H_{\mathrm{\mathrm{cl}}}^{(n)})$ w.r.t.\ $E$ increases with $nd$.

In the random case $V=V_{\omega}$ we obtain in place of
\eqref{cldensn}
\begin{equation}
\label{cldensnsto}
\begin{aligned}
\enn(E;H_{\mathrm{cl}}^{(n)})\;=\;&\frac{1}{nd}\mathrm{Vol}(\S^{nd-1})
(2m)^{nd/2}\\ & \cdot\int_\R d\mu(q)\int_{\Lambda_0^n}
\Theta(E-q\sum_{i}V_{0}(x_{i}))
(E-q\sum_{i}V_{0}(x_{i}))^{nd/2}d^{nd}x.
\end{aligned}
\end{equation}
For fixed number of particles $n$ the smoothness of
$\enn(E;H_{\mathrm{\mathrm{cl}}}^{(n)})$ depends on the probability distribution
$d\mu(q)$ of random variables $q_j(\omega)$. Assume that the measure $d\mu(q)$ is
absolutely continuous with smooth density supported in an interval
$[\epsilon,\epsilon^{-1}]$ for some $0<\epsilon<1$. Then we may rewrite
\eqref{cldensnsto} as
\begin{equation}
\label{cldensnsto1}
\begin{aligned}
\enn(E;H_{\mathrm{cl}}^{(n)})\;=\;&\frac{1}{nd}\mathrm{Vol}(\S^{nd-1})
(2m)^{nd/2}\\ & \cdot\int q^{nd/2}f(q)dq \int_{\Lambda_0}
\Theta\left(\frac{E}{q}-V_{0}(x)\right)
\left(\frac{E}{q}-V_{0}(x)\right)^{nd/2} d^{nd}x\\
=\;&E^{nd/2+1}\frac{1}{nd}\mathrm{Vol}(\S^{nd-1})
(2m)^{nd/2}\\ & \cdot\int q^{nd/2}f(Eq)dq \int_{\Lambda_0}
\Theta\left(\frac{1}{q}-V_{0}(x)\right)
\left(\frac{1}{q}-V_{0}(x)\right)^{nd/2}d^{nd}x.
\end{aligned}
\end{equation}
In particular, if $f\in C^\infty$ then the degree of differentiability of
$\enn(E;H_{\mathrm{cl}}^{(n)})$ is completely determined by the factor $E^{nd/2+1}$.

The main idea of the present work is to replace $N(E;H_{\mathrm{cl}}^{(n)})$ in the
relations \eqref{nc1} -- \eqref{t1} by the integrated density of states
$\enn(E;H_{\mathrm{cl}}^{(n)})$. For $nd\ge 3$ this gives the temperature
$T(E)=T(E;H_{\mathrm{cl}}^{(n)})$ as
\begin{eqnarray}
\label{TV}
\frac{1}{T(E)}&=
&\frac{d}{dE}\ln\frac{d}{dE}\enn(E;H_{\mathrm{cl}}^{(n)})\nonumber\\
&=&\left(\frac{nd}{2}-1\right)k\frac{\int_{\Lambda_0^n}
\Theta(E-V^{(n)}(x))(E-V^{(n)}(x))^{nd/2-2}d^{nd}x}{\int_{\Lambda_0^n}
\Theta(E-V^{(n)}(x))(E-V^{(n)}(x))^{nd/2-1}d^{nd}x},\quad nd\geq 3.
\end{eqnarray}
which is obviously nonnegative.

The motivation is as follows. We consider the periodic case only, the random case can
be treated similarly. For a given finite volume $\Lambda$ the temperature can be
defined by means of the relation \eqref{t1}, i.e.,
\begin{equation*}
\frac{1}{T(E;H_{\mathrm{cl},\Lambda}^{(n)})}=k\frac{d}{dE}
\ln \frac{dN(E;H^{(n)}_{cl,\Lambda})}{dE}
\end{equation*}
with
\begin{equation*}
N(E;H^{(n)}_{\mathrm{cl},\Lambda})={\int_{\Lambda^n}
\Theta(E-V^{(n)}(x))(E-V^{(n)}(x))^{nd/2-1}d^{nd}x}.
\end{equation*}
Considering the limit $\Lambda\rightarrow\infty$ in the Fisher sense due to the
relations
\begin{equation*}
\frac{d^k}{dE^k}\enn(E;H_{\mathrm{cl}}^{(n)})=
\lim_{\Lambda\rightarrow\infty}\frac{1}{|\Lambda|^n}
\frac{d^k}{dE^k}N(E;H^{(n)}_{cl,\Lambda}),\quad k=1,2,
\end{equation*}
we obtain $\displaystyle\lim_{\Lambda\rightarrow\infty}T(E;H^{(n)}_{cl,\Lambda})
=T(E;H_{\mathrm{cl}}^{(n)})$.

The condition
\begin{equation}
\label{ineqTV1}
\frac{d}{dE}\enn(E;H_{\mathrm{cl}}^{(n)})\frac{d^3}{dE^3}\enn(E;H_{\mathrm{cl}}^{(n)})
\le\left(\frac{d^2}{dE^2}\enn(E;H_{\mathrm{cl}}^{(n)})\right)^2,\quad nd\geq 5
\end{equation}
guarantees that the temperature $T(E;H_{\mathrm{cl}}^{(n)})$ is a non-decreasing
function of the energy, i.e., it naturally replaces condition \eqref{ineq1}. It may
be rewritten as
\begin{equation}
\label{ineqTV}
\left(\frac{nd}{2}-2\right)\langle{\cT_E}^{-2}\rangle_E
\leq\left(\frac{nd}{2}-1\right)\langle{\cT_E}^{-1}\rangle_E^2
\end{equation}
Here ${\cT_E(x)}^{-1}=(nd/2-1){(E-V^{(n)}(x))}^{-1}$ and $\langle\;\rangle_E$ is the
mean given by the probability measure
\begin{equation*}
d\mu_E(x)=\frac{\Theta(E-V^{(n)}(x))(E-V^{(n)}(x))^{nd/2-1}d^{nd}x}{\int_{\Lambda_0^n}
\Theta(E-V^{(n)}(x))(E-V^{(n)}(x))^{nd/2-1}d^{nd}x}
\end{equation*}
on $\Lambda_0^n$ provided $E$ is such that the set $\{x|V^{(n)}(x)\le E\}$ has
non-zero measure. Thus, we obtain the relation similar to
\eqref{tempav},
\begin{equation}
\label{meanper}
\frac{1}{T(E)}=
k\langle\,{\cT_E}^{-1}\rangle_E.
\end{equation}

Note that the heat capacity will be negative if the inequality opposite to
\eqref{ineqTV} holds. By the Schwarz inequality we have
\begin{equation}
\label{schwarz1}
\langle {\cT_E}^{-2}\rangle_E
\geq \langle {\cT_E}^{-1}\rangle_E^2.
\end{equation}
On the other hand $(nd/2-2)<(nd/2-1)$. Therefore for the heat capacity to be negative
the inequality \eqref{schwarz1} has to be sufficiently strict.

As an example where the temperature decreases with the energy consider the following
choice for the periodic potential. Let $X\subset\Lambda_0$ have measure $0<\alpha<1$.
Define $V(x)=E_0>0$ for $x\in X$ and zero otherwise when $x\in\Lambda_0\setminus X$.
Finally extend $V$ periodically to all of $\R^d$. In what follows $E_0$ will be
fixed. Then for any $E>nE_0$ and $k=1,2,3$
\begin{eqnarray}
\label{beispiel}
0\le\int_{\Lambda_0^n}
\Theta(E-V^{(n)}(x))(E-V^{(n)}(x))^{nd/2-k}d^{nd}x\nonumber\\
=\sum_{j=0}^n \binom{n}{j}\alpha^j(1-\alpha)^{n-j}(E-jE_0)^{nd/2-k}.
\end{eqnarray}
We now fix $\alpha$ to be given as $\alpha=1-((E-nE_0)/E)^{(nd/2-2)/n}$. Then the
leading contribution to
\eqref{beispiel} when $E-nE_0$ is small is given as
\begin{eqnarray*}
(E-nE_0)^{nd/2-3}& \mbox{   for }& k=3,\nonumber\\
\mathrm{const}\cdot(E-nE_0)^{nd/2-2}&
\mbox{   for }& k=1,2.
\end{eqnarray*}
When inserted into \eqref{ineqTV} the l.h.s.\ behaves to leading order as
$\mathrm{const}\cdot(E-nE_0)^{nd/2-5}$, whereas the r.h.s.\ behaves like
$\mathrm{const}\cdot(E-nE_0)^{nd/2-4}$ with positive constants depending on $E$, $n$
and $d$. Therefore when $E>nE_0$ is sufficiently close to $n E$ the inequality in
\eqref{ineqTV} is indeed reversed. Having thus fixed $\alpha$ by continuity
the heat capacity therefore also becomes negative for all energies sufficiently close
to $E$. Starting from the potential just constructed by a small change we may achieve
that $V$ is smooth. By the same arguments we may replace $\Lambda_0$ by any $\Lambda$
and find a potential $V$ supported in $\Lambda$ such that the heat capacity defined
by $N(E;H^{(n)}_{\mathrm{cl},\Lambda})$ is negative at least in some energy interval.

By the representation \eqref{meanper} we may also define the inverse temperature
fluctuation $\Delta(T^{-1})(E)=\Delta(T^{-1})(E;H^{(n)}_{\mathrm{cl}})$ by
\begin{equation}
\label{deltameanper}
\Delta(T^{-1})(E)^2=k^2\left(\langle\,{\cT_E}^{-2}\rangle_E-
\langle\,{\cT_E}^{-1}\rangle_E^2\right)\ge 0.
\end{equation}
Recalling the definition of the heat capacity as $1/c_{\mathrm{v}}(E)=dT(E)/dE$ we
obtain in the case $nd\geq 5$
\begin{equation*}
\frac{1}{c_{\mathrm{v}}(E)}=-kT(E)^2
\left[\frac{(\frac{nd}{2}-2)}{(\frac{nd}{2}-1)}
\langle\,{\cT_E}^{-2}\rangle_E-\langle\,{\cT_E}^{-1}\rangle_E^2\right]
\end{equation*}
and therefore
\begin{equation}
\label{heatvarcap1}
\Delta(T^{-1})(E)^2+\frac{k}{T(E)^2c_{\mathrm{v}}(E)}=
k^2\left(\frac{nd}{2}-1\right)^{-1}\langle\,{\cT_E}^{-2}\rangle_E\ge 0.
\end{equation}

In the absence of external fields, i.e., for $V=0$ this gives $\Delta(1/T)(E)=0$.
Since this is an undesirable feature, we will consider the following alternative. Let
$E^{(n)}_{\min} = \inf_{x\in\R^{nd}}V^{(n)}(x) >
-\infty$. Motivated by the construction \eqref{dnu} consider the
probability measure on the interval $[E^{(n)}_{\min},E]$
\begin{equation}
\label{dnu1}
d\nu_E(E^{\prime})=\frac{1}{\frac{d}{dE}\enn(E;H_{\mathrm{cl}}^{(n)})}
\frac{d^2}{dE^{\prime^2}} \enn(E^{\prime};H_{\mathrm{cl}}^{(n)})\;
dE^{\prime},
\end{equation}
such that all derivatives (provided they exist) of
$\enn(E^{\prime};H_{\mathrm{cl}}^{(n)})$ with respect to $E^{\prime}>E^{(n)}_{\min}$
are positive. Denote the resulting mean by $\langle\;\rangle_E^{\prime}$. Observe
that
\begin{equation*}
\left.\frac{d\nu_E(E^{\prime})}{dE^{\prime}}\right|_{E^{\prime}=E-0}
=\frac{1}{kT(E)}
\end{equation*}
and so again we have
\begin{equation*}
\frac{1}{T(E)}=k\langle {\cT}^{-1}\rangle^{\prime}_E
\end{equation*}
with
\begin{equation*}
{\cT(E^{\prime})}^{-1}
=\frac{d}{dE^{\prime}}\ln\frac{d^2}{dE^{\prime^2}}
\enn(E^{\prime};H_{\mathrm{cl}}^{(n)}).
\end{equation*}
In analogy to \eqref{flucprime} this results in an alternative definition of the
fluctuation of the inverse temperature of the form
\begin{equation*}
\Delta^{\prime}(T^{-1})(E)=
\left(k^2\langle {\cT}^{-2}\rangle_E^{\prime}
-T(E)^{-2}\right)^{1/2}
\end{equation*}
and which in general differs from $\Delta(T^{-1})(E)$ given by
\eqref{deltameanper}. For comparison we first observe that $(nd\ge 5)$
\begin{equation*}
\langle{\cT_E}^{-2}\rangle_E=
\frac{(\frac{nd}{2}-1)}{(\frac{nd}{2}-2)}
\frac{d^3}{dE^{3}} \enn(E;H_{\mathrm{cl}}^{(n)})
\bigg/\frac{d}{dE} \enn(E;H_{\mathrm{cl}}^{(n)}).
\end{equation*}
On the other hand
\begin{equation}
\label{comp1}
\langle{\cT}^{-2}\rangle_E^{\prime}=
\int_{E^{(n)}_{\min}}^{E}
\left(\left(\frac{d^3}{dE^{\prime^{\,3}}} \enn(E^{\prime};H_{\mathrm{cl}}^{(n)})\right)^2
\bigg/\frac{d^2}{dE^{\prime^{\,2}}} \enn(E^{\prime};H_{\mathrm{cl}}^{(n)})\right)
dE^{\prime}\bigg/\frac{d}{dE} \enn(E;H_{\mathrm{cl}}^{(n)}).
\end{equation}

For $nd\ge 7$ we perform a partial integration in \eqref{comp1} and obtain by
comparison
\begin{equation}
\label{comp2}
\langle{\cT}^{-2}\rangle_E^{\prime}=
\frac{(\frac{nd}{2}-2)}{(\frac{nd}{2}-1)}\langle {\cT_E}^{-2}\rangle_E
+\langle\tilde{\cG}\rangle_E^{\prime}
\end{equation}
with
\begin{equation*}
\tilde{\cG}(E^{\prime})=-\frac{d}{dE^{\prime}}\frac{1}{\cT(E^{\prime})}.
\end{equation*}
The condition $\cG(E^{\prime})\ge 0$, which means that $\cT(E^{\prime})$ increases at
$E^{\prime}$, is equivalent to
\begin{equation*}
\frac{d^2}{dE^{\prime^{\,2}}}\enn(E^{\prime};H_{\mathrm{cl}}^{(n)})
\frac{d^4}{dE^{\prime^{\,^4}}}\enn(E^{\prime};H_{\mathrm{cl}}^{(n)})
\le\left(\frac{d^3}{dE^{\prime^{\,^3}}}\enn(E^{\prime};H_{\mathrm{cl}}^{(n)})
\right)^2,
\end{equation*}
which compares with \eqref{ineqTV1} and hence may be discussed in a similar way. By
\eqref{heatvarcap1} and \eqref{comp2} the heat capacity can be expressed as
\begin{equation*}
\frac{1}{c_{\mathrm{v}}(E)}=-k^2\left[\langle{\cT}^{-2}\rangle_E^{\prime}
-\langle\tilde{\cG}\rangle_E^{\prime}
-\langle {\cT}^{-1}\rangle_E^{\prime{\,^2}}\right]
\end{equation*}
This alternative definition of the fluctuation and the heat capacity are therefore
related by
\begin{equation}
\label{comp11}
\Delta^{\prime}(T^{-1})(E)^2+\frac{k}{T(E)^2c_{\mathrm{v}}(E)}=kT(E)^2
\langle\tilde{\cG}\rangle_E^{\prime}.
\end{equation}

As a test in the special case $V=0$, i.e., in the absence of external fields, and for
$nd>6$ this new fluctuation is calculated explicitly to be
\begin{equation}
\label{deltafree}
\Delta^{\prime}(T^{-1})(E)= \left(\frac{nd}{2}-3\right)^{-1/2}
\left(\frac{nd}{2}-1\right)^{-1/2}
\frac{1}{T(E)}.
\end{equation}
If we {\it define} the temperature fluctuation as
$\Delta^{\prime}(T)(E)=\Delta^{\prime}(T^{-1})(E)\,T(E)^2$ then for large $nd$ we
obtain $\Delta^{\prime}(T)(E)\approx (2/nd)T(E)$, now as it should be.

The following asymptotic expansion holds for large $E$ and $nd>2$
\begin{equation}\label{asymTV}
\begin{split}
\frac{\Gamma(\frac{nd}{2}+1)}{(2\pi m)^{nd/2}}
\frac{d}{dE}\enn(E;H_{\mathrm{cl}}^{(n)}) & =
\frac{nd}{2}\int_{\Lambda_0^n}\Theta(E-V^{(n)}(x))
(E-V^{(n)}(x))^{nd/2-1}d^{nd}x\\ &= \frac{nd}{2}E^{nd/2-1} \left[1- a E^{-1} + b
E^{-2} +O(E^{-3})\right],
\end{split}
\end{equation}
where
\begin{equation*}
\begin{split}
a & = \left(\frac{nd}{2}-1\right) \int_{\Lambda_0^n}V^{(n)}(x)d^{nd}x
= n\left(\frac{nd}{2}-1\right)
\overline{V},\\
b &= \frac{1}{2}\left(\frac{nd}{2}-1\right)\left(\frac{nd}{2}-2\right)
\int_{\Lambda_0^n}(V^{(n)}(x))^2d^{nd}x\\ & = \frac{1}{2}
\left(\frac{nd}{2}-2\right)\left(\frac{nd}{2}-1\right)
\left[n\overline{V^2}+n(n-1)\overline{V}^2\right].
\end{split}
\end{equation*}
Here we use the following notation. In the periodic case and for any natural number
$k$
\begin{equation}
\label{Vav}
\overline{V^k}=\int_{\Lambda_0}V(x)^kd^{d}x=\int_{\Lambda_0}V_{0}(x)^kd^{d}x.
\end{equation}
In the stochastic case
\begin{equation}
\label{Vavsto}
\overline{V^k}=\overline{q^k}\int_{\Lambda_0}V_{0}(x)^kd^{d}x\quad\text{with}\quad
\overline{q^k}=\int q^{k}d\mu(q).
\end{equation}
Note that $\overline{V}^2\le\overline{V^2}$ by the Schwarz
inequality with equality if and only if $V$ is constant in $x$ (and in
$\omega$ in the random case). This gives the following
asymptotic energy-temperature relation
\begin{equation}
\label{asymTV1}
\begin{aligned}
\frac{1}{T(E;H_{\mathrm{cl}}^{(n)})}\:=\:&
k\left(\frac{nd}{2}-1\right)E^{-1}\\ &\cdot\Big(1+E^{-1}n\overline{V}
+E^{-2}n\left[\left(n+\frac{nd}{2}-2\right)\overline{V}^2
-\left(\frac{nd}{2}-2\right)\overline{V^2}\right]
+O(E^{-3})\Big).
\end{aligned}
\end{equation}
Therefore for all sufficiently large $E$ and for $nd>\,2$ the temperature increases
with $E$. Also when $nd>\,2$ at fixed large energy $E$ to leading
order in $E^{-1}$ the temperature $T(E)$ decreases (as a function of the potential) when a potential
with $\overline{V}>\,0$ is switched on. Otherwise it decreases. Similarly if
$\overline{V}=0$ such that $V$ is not a constant, then $T(E)$ decreases for all large
$nd$ when such a potential is switched on.

For the heat capacity we obtain the asymptotic expansion for large
energies
\begin{equation}
\label{asymTV2}
c_{\mathrm{v}}(E;H_{\mathrm{cl}}^{(n)})=k\left(\frac{nd}{2}-1\right)
\left(1+E^{-2}n\left(\frac{nd}{2}-2\right)
\left[\overline{V^2}-\overline{V}^2\right]+O(E^{-3})\right).
\end{equation}
Thus for large $E$ and $nd>4$ the heat capacity increases when a potential is
switched on.

A similar discussion for a canonical ensemble is possible by defining the partition
function as the Laplace-Stieltjes transform of the integrated density of states
\begin{eqnarray*}
z(\beta;H_{\mathrm{cl}}^{(n)})&=&\int e^{-\beta E} d\enn(E;H_{\mathrm{cl}}^{(n)})
=\beta \int e^{-\beta E}\enn(E;H_{\mathrm{cl}}^{(n)})dE\nonumber\\
&=&\lim_{\Lambda\rightarrow\infty}\frac{1}{|\Lambda|^{n}}
\int e^{-\beta E}dN(E;H_{cl,\Lambda}^{(n)}),
\end{eqnarray*}
such that $z(\beta;H_{0\,\mathrm{cl}}^{(n)})=(2\pi mkT)^{nd/2}$. This limit has not
to be confused with the notion of ``thermodynamic limit'' used when discussing
extensive systems in statistical mechanics. There one divides the logarithm of the
partition function in a finite volume by the volume of the configuration space and
then takes the limit (see, e.g., \cite{Rue}). So in order to distinguish these two
limits we use the notion ``per unit interaction volume'' (see,
however, the discussion of the quantum mechanical grand canonical ensemble in Section
\ref{sec:3}). The reason for this difference is as follows. The thermodynamic limit
has to be taken when the number of particles increases with
$\Lambda$ such that the density remains constant. Here the system is not extensive
since the number of particles stays fixed.

The resulting entropy, mean energy and the Helmholtz free energy are then given as
\begin {equation*}
\begin{split}
s(\beta;H_{\mathrm{cl}}^{(n)})\ & =\ -k\beta^2\frac{d}{d\beta}
(\frac{1}{\beta}z(\beta;H_{\mathrm{cl}}^{(n)})),\\
\bar{e}(\beta;H_{\mathrm{cl}}^{(n)})\ & =\ -\frac{d}{d\beta}
\ln z(\beta;H_{\mathrm{cl}}^{(n)}),\\
f(\beta;H_{\mathrm{cl}}^{(n)})\ & =\ -\frac{1}{\beta}\ln z(\beta;
H_{\mathrm{cl}}^{(n)}),
\end{split}
\end{equation*}
giving the familiar relation between these three quantities
\begin{equation}
\label{meef1}
\bar{e}(\beta;H_{\mathrm{cl}}^{(n)})=Ts(\beta;H_{\mathrm{cl}}^{(n)})+
f(\beta;H_{\mathrm{cl}}^{(n)}).
\end{equation}
In analogy to \eqref{canheatcap} the heat capacity in this canonical setup is given
as $c_{\mathrm{v}}(T;H_{\mathrm{cl}}^{(n)})
= d\bar{e}(\beta;H_{\mathrm{cl}}^{(n)})/dT$.

Also in analogy to the discussion following \eqref{ccan1} we may relate the
microcanonical and the canonical description. We will assume $nd>5$ such that
$\enn(E;H_{\mathrm{cl}}^{(n)})$ is three times differentiable w.r.t. $E$. Write
\begin{equation}
\label{cg}
z(\beta;H_{\mathrm{cl}}^{(n)})=\int e^{-\beta E}\frac{d}{dE}
\enn(E;H_{\mathrm{cl}}^{(n)})dE
  =\beta\int e^{-\beta E
     +\ln(\frac{1}{\beta} \frac{d}{dE}\enn(E;H_{\mathrm{cl}}^{(n)}))}dE.
\end{equation}
Then the integrand takes its extremal value at $E_{\max}=E_{\max}(T)$ given
implicitly by
\begin{equation}
\label{T2}
\frac{1}{T}=k\frac{d}{dE}\ln\frac{d}{dE}\enn(E;H_{\mathrm{cl}}^{(n)})
|_{E=E_{\max}}.
\end{equation}
Relation \eqref{T2} now compares with \eqref{TV}. In other words the temperature
defined in the microcanonical ensemble at the energy $E=E_{\max}$ agrees with the
chosen one for the canonical ensemble. If a solution $E_{\max}$ to this relation
exists it is unique  and a maximum provided the inequality \eqref{ineqTV} holds for
all $E$ (for all large $E$ this is true by \eqref{asymTV2}) and is a strict
inequality at $E=E_{\max}$. Indeed, for $E$ near $E_{\max}$ we obtain
\begin{equation*}
-\beta E +\ln(\frac{1}{\beta}
 \frac{d}{dE}\enn(E;H_{\mathrm{cl}}^{(n)}))=-\frac{1}{2}\alpha(E-E_{\max})^2
+O((E-E_{\max})^3)
\end{equation*}
with $\alpha=\alpha(\beta;H_{\mathrm{cl}}^{(n)})$ now given by
\begin{equation*}
\alpha=\left.-\frac{d^2}{dE^2}\ln\frac{d}{dE}\enn(E;H_{\mathrm{cl}}^{(n)})
\right|_{E=E_{\max}},
\end{equation*}
which is nonnegative if \eqref{ineqTV} holds and is positive if
\eqref{ineqTV} is a strict inequality at $E=E_{\max}$. Again as in the discussion
after \eqref{ccan1} and in the microcanonical notation, by \eqref{T2} $\alpha$ is
related to the heat capacity via $\alpha=1/(k
c_{\mathrm{v}}(E_{\max})T(E_{\max})^2)$.

Hence to a good approximation and for given temperature the canonical ensemble is
described in terms of a Gaussian distribution in energy with variance $\alpha^{-1/2}$
in case $\alpha$ is positive and large. In particular in this approximation we obtain
$\bar{e}(\beta;H_{\mathrm{cl}}^{(n)})\approx E_{\max}$ and
$\Delta\bar{e}(\beta;H_{\mathrm{cl}}^{(n)})\approx\alpha^{-1/2}$. Similarly the
canonical heat capacity and the microcanonical heat capacity at energy $E_{\max}$
agree in this approximation (see the discussion after \eqref{ccan4}). Observe that
for large $T$ the implicit equation \eqref{T2} has always at least one solution
$E_{\max}\approx (nd/2-1)kT$ due to \eqref{asymTV}, which by \eqref{asymTV2} is at
least a local maximum. In the free case by \eqref{T2} $E_{\max}=(nd/2-1)kT$ (which
compares with the energy-temperature relation $E=(nd/2-1)T(E)$ in the microcanonical
ensemble for the free case) and $\alpha^{-1}=(nd/2-1)^{-1}E_{\max}^2=(nd/2-1)(kT)^2$.
So as a function of the particle number $n$ in the free case the variance
$\alpha^{-1/2}$ goes like $n^{-1/2}$ when the energy is fixed and like $n^{1/2}$ when
the temperature is fixed. We expect this feature to extend to the general case. This
would in particular imply that at fixed given energy the difference between the
microcanonical and the canonical description decreases with increasing particle
number.

For periodic and bounded potentials the following asymptotic expansion holds for
small $\beta$
\begin{eqnarray}
\label{asymeef}
z(\beta;H_{\mathrm{cl}}^{(n)})&=&z(\beta;H_{0\,\mathrm{cl}}^{(n)})
                \bigg(1-\beta n\overline{V}
+\frac{\beta^2}{2}\left[n\overline{V^2}+n(n-1)\overline{V}^2\right]
+O(\beta^3)\bigg).
\end{eqnarray}
Note that the Laplace transform of the asymptotic expansion \eqref{asymTV} gives this
asymptotic expansion. Relation \eqref{asymeef} results in the following asymptotic
temperature-energy relation
\begin{equation}\label{mteasy}
\bar{e}(\beta;H_{\mathrm{cl}}^{(n)})=\frac{nd}{2}\beta^{-1}+
n\overline{V}-\beta n
\left[\overline{V^2}-\overline{V}^2\right]+O(\beta^2).
\end{equation}
So for fixed small $\beta$ to leading order the mean energy increases when a
potential with $\overline{V}>0$ is switched on. Otherwise it decreases.  This
behavior is in agreement with the corresponding result \eqref{asymTV1} in the
microcanonical description. If $\overline{V}=0$ then the mean energy increases when
such a potential is switched on.

For large $E$ and $T$ respectively the comparison between the microcanonical and the
canonical description can be made more explicit. Indeed, relation \eqref{mteasy} may
be inverted to give
\begin{eqnarray}
\label{tmeasy}
\frac{1}{T}&=
&k\frac{nd}{2}\bar{e}(\beta;H_{\mathrm{cl}}^{(n)})^{-1}
\bigg(1+\bar{e}(\beta;H_{\mathrm{cl}}^{(n)})^{-1}n\overline{V}\nonumber\\
&&+\bar{e}(\beta;H_{\mathrm{cl}}^{(n)})^{-2}n\left[\left(n+\frac{nd}{2}\right)\overline{V}^2
-\frac{nd}{2}\overline{V^2}\right]
+O(\bar{e}(\beta;H_{\mathrm{cl}}^{(n)})^{-3})\bigg)
\end{eqnarray}
which compares with the microcanonical relation \eqref{asymTV1}. In other words for
$E$ and $n$ large the temperature defined in the microcanonical ensemble agrees with
the one chosen for the canonical ensemble, i.e., we have
\begin{equation}
\label{clinv}
T\approx T(\bar{e}(\beta=1/kT;H_{\mathrm{cl}}^{(n)})),\qquad E\approx
\bar{e}(\beta=1/kT(E;H_{\mathrm{cl}}^{(n)});H_{\mathrm{cl}}^{(n)}).
\end{equation}

{}From \eqref{mteasy} we obtain the asymptotic expansion for small
$\beta$
\begin{equation*}
c_{\mathrm{v}}(\beta;H_{\mathrm{cl}}^{(n)})=\frac{knd}{2}+\frac{kn\beta^2}{2}
\left[\overline{V^2}-\overline{V}^2\right]+O(\beta^3).
\end{equation*}
So if a potential is switched on, the heat capacity increases when $\beta$ is small.
Setting
\begin{equation}
\label{varebar}
\Delta\bar{e}(\beta;H_{\mathrm{cl}}^{(n)})=
\left(\frac{d^2}{d\beta^2}\ln z(\beta;H_{\mathrm{cl}}^{(n)})\right)^{1/2}
=(kT^2c_{\mathrm{v}}(T;H_{\mathrm{cl}}^{(n)}))^{1/2}
\end{equation}
to be the energy fluctuation we obtain for its asymptotic behavior
\begin{equation}
\label{ebarascl}
\Delta\bar{e}(\beta;H_{\mathrm{cl}}^{(n)})
=\sqrt{\frac{nd}{2}}\frac{1}{\beta}
\left(1+\frac{\beta^2}{d}\left[\overline{V^2}
-\overline{V}^2\right]+O(\beta^3)\right).
\end{equation}

\section{The quantum theory}\label{sec:3}
\setcounter{equation}{0}

In analogy to the classical case the microcanonical ensemble for a quantum system is
usually given as follows. Let $\{\cH,H\}$ be a quantum mechanical
system, where $\cH$ is a Hilbert space and $H$ a Hamilton
operator on $\cH$ with a purely discrete spectrum. Motivated by the classical case on
might be tempted to define the the following microcanonical quantity
\begin{eqnarray*}
\frac{d}{dE}N(E;H)&=&\frac{d}{dE}\tr_{\cH}(\Theta(E-H))\quad =\quad \tr_{\cH}(\delta(E-H)).
\end{eqnarray*}
As is well known (see, e.g., \cite{KHuang}) this procedure encounters difficulties
for the following reason. Since $H$ has a purely discrete
spectrum $N(E;H)=\tr_{\cH}(\Theta(E-H))$ is given as the number of eigenvalues
(counting multiplicities) of $H$ up to energy $E$. But then $dN(E;H)/dE$ is a sum of
$\delta$-functions at the eigenvalues of $H$ with coefficients given by the
multiplicities of these eigenvalues. In other words $dN(E;H)/dE$ is a
generalized function. On the other hand if $H$ has continuous spectrum
then both $\tr_{\cH}(\Theta(E-H))$ and $\tr_{\cH}(\delta(E-H))$ do not make sense.
This is in our point of view the main obstacle in making the microcanonical concept
useful in the quantum context. The usual way out is to consider ``large systems'',
where the spacing of the eigenvalues becomes small and to replace $dN(E;H)/dE$ by
$\tr_{\cH} (\Theta(E+\Delta E-H)-\Theta(E-H))$ with $\Delta E$ being small. This,
however, leads to another dilemma, since usually there is no natural and intrinsic
choice for the size of $\Delta E$.

The approach we will propose will give a precise meaning to the notion ``large
system'' in the sense that the number of particles $n$ only
should be sufficiently large. The one-particle operators $H=H^{(1)}$ we will consider
are supposed to have continuous spectrum. More precisely, we will consider
one-particle Schr\"{o}dinger operators acting on the Hilbert space $L^2(\R^d)$ and of the
form
\begin{equation*}
H=H_0 +V,
\end{equation*}
Here $H_0$ is the free Hamiltonian for a particle of mass $m>0$
\begin{equation*}
H_0=-\frac{\hbar^2}{2m}\Delta
\end{equation*}
where $\Delta$ is the Laplace operator on $\R^d$, the configuration space for the
particle. The potentials $V$ we have in mind are periodic or random as in
\eqref{stopo}.

The resulting $n$-particle operators with no interaction between the
particles will be denoted by $H^{(n)}$. They act on the $n$-fold
tensor product $\otimes_n L^2(\R^d)$ of $L^2(\R^d)$, which is equal to
$L^2(\R^{nd})$, if we consider the particles to be distinguishable or on the
subspaces $\otimes_{n,\pm} L^2(\R^d)$ of symmetric $(+)$ or antisymmetric $(-)$ wave
functions, when we allow for statistics.

Let $\chi_{\Lambda}$ be the characteristic function of any cube $\Lambda\subset \R^d$
and correspondingly let $\chi_{\Lambda^n}$ denote the
characteristic function of $\Lambda^n\subset\R^{nd}$. Set $V_{\Lambda}
= \chi_{\Lambda}V$. Also let $\Delta_{\Lambda}$ be the Laplace operator on $\Lambda$
with Dirichlet or Neumann boundary conditions. Let
$H_{\Lambda}=H_{0,\Lambda}+V_{\Lambda}$ with
$H_{0,\Lambda}=-\hbar^2/2m\cdot\Delta_{\Lambda}$ and correspondingly
$H_{\Lambda}^{(n)}$ for the resulting $n$-particle operator. In the examples we have
in mind all $H_{\Lambda}^{(n)}$ have a discrete spectrum with no finite accumulation
points such that $N(E;H_{\Lambda}^{(n)})$ as well as the canonical Boltzmann-Gibbs
partition function is well defined. Furthermore it can be shown that the limit,
called the integrated density of states,
\begin{equation}
\label{dens0}
0\le \enn(E;H^{(n)})=\lim_{\Lambda\rightarrow \infty}
\frac{1}{|\Lambda|^n}N(E;H_{\Lambda}^{(n)})
\end{equation}
exists \cite{Pastur:Figotin} and is independent of the boundary conditions \cite{Na}.
In particular in the free $n$-particle case $\enn(E;H^{(n)}_{0})=0$ for $E<0$
\begin{equation}
\label{densn0}
\enn(E;H^{(n)}_0)=\frac{1}{\Gamma(\frac{nd}{2}+1)}
\left(\frac{mE}{2\pi\hbar^2}\right)^{nd/2}
=\frac{(2\pi mE)^{nd/2}}{h^{nd}\Gamma(\frac{nd}{2}+1)}
\end{equation}
for $E>0$. This reflects the well known intuitive observation
that each quantum state occupies the volume $h^n\;(h=2\pi\hbar)$ in the phase space
$\R^{2nd}$. In other words a comparison with the classical case \eqref{cldens0} gives
the relation $h^{nd}\enn(E;H^{(n)}_0)=\enn(E;H^{(n)}_{0\,\mathrm{cl}})$. So also
$\enn(E;H^{(n)}_0)$ is smooth for $E\neq 0$ and $[nd/2]$ times differentiable at
$E=0$ (here $[a]$ denotes integer part of $a$). The smoothness of
$\enn(E;H^{(n)}_0)$, i.e., the order of differentiability,
increases with the number of particles.

For random potentials, i.e., when we consider $H=H_{\omega}=H_0+V_{\omega}$, the
integrated density of states $\enn(E;H_{\omega})$ is actually a deterministic
quantity, i.e., independent of $\omega$ for almost all $\omega$. The same is also
true for the spectrum $\spec(H)$ of $H_{\omega}$ and in particular for its infimum
$\inf\spec (H_{\omega})$ \cite{Pastur:Figotin}. This will have the important
consequence that all other quantities we will introduce using the integrated density
of states will also be deterministic. So in order to cover both
periodic and random potentials simultaneously we will simply write $\enn(E;H)$ for
$\enn(E;H_{\omega})$ and similarly $\enn(E;H^{(n)})$ for the $n$-particle case.

Obviously $\enn(E;H^{(n)})$ is a monotone increasing function in $E$
and therefore
continuous at almost all $E$. Its derivative, if it exists, is called the density of
states for $H^{(n)}$. Also by construction $\enn(E+c;H^{(n)}+c)=\enn(E;H^{(n)})$ for
any constant $c$ and $n(\lambda E;\lambda
H^{(n)})=\enn(E;H^{(n)})$ for all $\lambda
>0$. Note also that in general $\enn(E;H^{(n)})$ need not vanish for $E<0$.

If the potential is sign-definite, i.e., $\pm V\ge 0$ such that
$\pm V_{\Lambda}\ge 0$ for all $\Lambda$ then by the min-max principle (see, e.g.,
\cite{CH1}) we have $\mp N(E;H_{\Lambda}^{(n)})\ge \mp N(E;H_{0,\Lambda})$ for all
$\Lambda$ and $E$ and $n$. This gives
\begin{equation}
\label{pm}
\mp \enn(E;H^{(n)})\ge\mp \enn(E;H_0^{(n)}) \mbox{  for all E if }\;
\pm V\ge 0
\end{equation}
and more generally
\begin{equation}
\label{pm1}
\enn(E;(H_0+V_{1})^{(n)})\le \enn(E;(H_0+V_{2})^{(n)})\mbox{  for all E if }\;
V_{1}\ge V_{2}.
\end{equation}

Alternatively the integrated density of states may be written
in the form
\begin{equation}
\label{densn}
\enn(E;H^{(n)})=\lim_{\Lambda\rightarrow \infty}
\frac{1}{|\Lambda|^n}\tr_{L^2(\R^{nd})}
(\chi_{\Lambda^n}\Theta(E-H^{(n)})).
\end{equation}
Therefore, for periodic $V$
\begin{equation}
\label{densn1}
\enn(E;H^{(n)})=\tr_{L^2(\R^{nd})}(\chi_{\Lambda^n_0}\Theta(E-H^{(n)})),
\end{equation}
where $\Lambda_0$ is the unit cube in $\R^d$. This relation also continues to hold
for random operators, when the right hand side is replaced by its average over the
random variable. That the integrated density of states is actually deterministic is
due to the limit relation in \eqref{densn}, which is called a self-averaging effect.
Below we will show how to obtain $\enn(E;H^{(n)})$ for all particle numbers $n$ from
the one particle integrated density of states $\enn(E;H)$.

One-particle scattering theory may now be used to relate $\enn(E;H_0+V)$ and
$\enn(E;H_0)$. More precisely let $\delta(E;H_0+V_{\Lambda},H_0)$ denote the phase
shift at energy $E>0$ for the pair $(H_0+V_{\Lambda},H_0)$. In
other words, if $S(E;H_0+V_{\Lambda},H_0)$ is the scattering matrix for the pair
$(H_0+V_{\Lambda},H_0)$ at energy $E$ then $\det\, S(E;H_0+V_{\Lambda},H_0)=\exp(2
i\delta(E;H_0+V_{\Lambda},H_0))$. For $E>0$ one has \cite{KS0,KS1}
\begin{eqnarray}
\label{dens1}
\enn(E;H)-\enn(E;H_0)=\lim_{\Lambda\rightarrow \infty}
\frac{1}{\pi |\Lambda|}\delta(E;H_0+V_{\Lambda},H_0)
=\frac{1}{\pi}\hat{\delta}(E;H,H_0).
\end{eqnarray}
Now $-\delta(E;H_0+V_{\Lambda},H_0)/\pi$ may be replaced by the so-called spectral
shift function $\xi(E;H_0+V_{\Lambda},H_0)$ of Krein (see, e.g.,
\cite{Birman:Yafaev}) which for negative $E$ equals minus the number
of bound states below $E$. We use the spectral shift function to extend the
scattering phase to all energies in such a way that
$\delta(E;H_0+V_{\Lambda},H_0)=-\pi \xi(E;H_0+V_{\Lambda},H_0)$. With this extension
of $\delta(E;H_0+V_{\Lambda},H_0)$ to all $E$ relation \eqref{dens1} also extends to
all $E$. In particular, for random potentials $\hat{\delta}(E;H,H_0)$ is
deterministic and $\enn(E;H)=\hat{\delta}(E;H,H_0)=0$ for $E<\min (\inf\spec(H),0)$.

The trace formula for the spectral shift function leads to the
following relation which is due to Beth and Uhlenbeck \cite{BU},
\begin{equation}
\label{Krein}
\tr_{L^2(\R^d)}(e^{-\beta (H_0+V_{\Lambda})}-e^{-\beta H_0})
=\frac{\beta}{\pi}\int e^{-\beta E}\delta(E;H_0+V_{\Lambda},H_0)dE,\qquad \beta >0.
\end{equation}

The spectral shift density $\hat{\delta}(E;H,H_0)$ in
\eqref{dens1} may be interpreted as the phase shift at energy $E$ per unit
interaction volume. In short we will call it the (total) scattering phase density.
Also \eqref{dens1} extends to the $n$-particle case giving
\begin{equation}
\label{dens2}
\enn(E;H^{(n)})-\enn(E;H_0^{(n)})=\lim_{\Lambda\rightarrow \infty}
\frac{1}{\pi |\Lambda|^n}\delta(E;H_0^{(n)}+V_{\Lambda^n}^{(n)},H_0^{(n)})
=\frac{1}{\pi}\hat{\delta}(E;H^{(n)},H_0^{(n)}).
\end{equation}
Here we have set $V_{\Lambda^n}^{(n)}(x)
=\chi_{\Lambda^n}(x)V^{(n)}(x)$ in analogy to the definition of $V_{\Lambda}$.
Also $\enn(E;H^{(n)})=\hat{\delta}(E;H^{(n)},H^{(n)}_0)=0$ for $E<n\min
(\inf\spec(H),0)$.

A precursor to \eqref{dens1} due to Friedel \cite{Fr1,Fr2} (see also
\cite{AM,Faulkner,Newton:91,M}) in the context of a single impurity is of course well
known in solid state physics. To the best of our knowledge relation
\eqref{dens1} and its extension \eqref{dens2} to the integrated density of states
seems to be new. Note that in case of two impurities described by two potentials
$W_1$ and $W_2$, say of finite range and even non-overlapping, one has
$\delta(E;H_0+W_1+W_2,H_0)\neq
\delta(E;H_0+W_1,H_0)+\delta(E;H_0+W_2,H_0)$. Therefore the
contribution from several impurities is not simply the sum of the contributions from
the individual impurities. However, contributions from different impurities are
asymptotically additive, when the distance between them becomes large
\cite{cluster1,cluster2}.

Relations \eqref{dens1} and \eqref{Krein} are in accordance with the well known chain
rule for the phase shift
\begin{equation}
\label{chain}
\delta(E;H_1,H_2)+\delta(E;H_2,H_3)=\delta(E;H_1,H_3),
\end{equation}
valid for any triple of Hamilton operators for which the scattering
phase for any of the pairs exist. In particular, the relation \eqref{chain}
implies $\delta(E;H_1,H_2)=-\delta(E;H_2,H_1)$. Also the equalities
\begin{equation}
\label{rules}
\delta(\lambda E;\lambda H_1,\lambda H_2)=\delta(E+c;H_1+c,H_2+c)
=\delta(E;H_1,H_2)
\end{equation}
hold for any $\lambda>0$ and any real $c$. Relations \eqref{rules} obviously extend
to $\hat{\delta}(E;H,H_0)$. The monotonicity property
\eqref{pm} is related to the well known relation for the scattering phase
\begin{equation*}
\mp\delta(E;H_0+V_{\Lambda},H_0) \ge 0 \mbox{ for all E if }\;
\pm V\geq 0,
\end{equation*}
and more generally (compare \eqref{pm1})
\begin{equation*}
\delta(E;H_0+V_{1,\Lambda},H_0)\le \delta(E;H_0+V_{2,\Lambda},H_0)
\mbox{ for all E if }\;
V_{1}\geq V_{2}.
\end{equation*}

If $V$ has compact support, then $\delta(E;H_0+V_{\Lambda},H_0)$ does
not depend on $\Lambda$ for all large $\Lambda$ and hence in that
case $\enn(E;H)=\enn(E;H_0)$. Below (see \eqref{conv2}) we will see
that this implies
$\enn(E;H^{(n)})
= \enn(E;H_0^{(n)})$ for all $n$. As mentioned in the introduction it would be
interesting to obtain information from $n$-particle scattering theory on
$\enn(E;H_0^{(n)} + V^{(n)}+W^{(n)})$ where the $n$-particle interaction $W^{(n)}$ is
given in terms of two-particle interaction potentials.

{}From the discussion so far it is clear that the quantity $\enn(E;H^{(n)})$ is a
nice quantum candidate to replace the distribution $\tr_{\cH}\Theta(E-H)$ which
replaces the classical distribution $N(E;H_{\mathrm{cl}})$. In particular if
$\enn(E;H^{(n)})$ is differentiable such that necessarily $d\enn(E;H^{(n)})/dE \ge 0$
then we may define
\begin{equation}
\label{sm}
s(E;H^{(n)})=k\ln \frac{d\enn(E;H^{(n)})}{dE}
\end{equation}
to be the entropy per unit (interaction) volume for a system of $n$ particles. We
hasten to point out that this definition of the entropy needs a choice of an energy
unit since $d\enn(E;H^{(n)})/dE$ has the dimension of an inverse energy. In other
words, $s(E;H^{(n)})$ is only well defined up to a constant. One way out is to
\emph{renormalize} the entropy additively by choosing a given energy $E_0$ and to
replace $s(E;H^{(n)})$ by $s(E;H^{(n)})-s(E_0;H^{(n)})$. Another way is to consider a
relative entropy (see, e.g.,  \cite{W} for a general discussion) say for the pair
$(H^{(n)},H^{(n)}_0)$ as $s(E;H^{(n)},H^{(n)}_0) =s(E;H^{(n)})-s(E;H^{(n)}_0)$.

In case $\enn(E;H^{(n)})$ is even twice differentiable w.r.t. $E$ a temperature
$T(E)=T(E;H^{(n)})$ may be defined by
\begin{equation}
\label{T1}
\frac{1}{T(E)}=\frac{d}{dE}s(E;H^{(n)})=
k\frac{d^2 \enn(E;H^{(n)})/d^2E}{d\enn(E;H^{(n)})/dE}
\end{equation}
which is independent of the choice of the energy unit. Obviously $T(E)$ is positive
if the second derivative of $\enn(E;H^{(n)})$ is positive and below we will show that
this is indeed the case for the Hamiltonians we consider if we restrict the particle
number $n$ to be $\ge 4$.

In analogy to the classical case (see \eqref{ineq1}) the temperature
$T(E)$ defined by \eqref{T1} is a monotone
increasing function at $E$ if $\enn(E;H^{(n)})$ is three times
differentiable and satisfies the inequality
\begin{equation*}
\frac{d^3}{dE^3}\enn(E;H^{(n)})\cdot \frac{d}{dE}\enn(E;H^{(n)})\le
\left(\frac{d^2}{dE^2}\enn(E;H^{(n)})\right)^2.
\end{equation*}
In this case the heat capacity $c_{\mathrm{v}}(E) $ defined
again by $1/c_{\mathrm{v}}(E) = dT(E)/dE$ in this quantum mechanical, microcanonical
context is positive. In the opposite case, i.e., with the inequality reversed, it
will be negative.

As announced in the previous section the fluctuation of the inverse temperature may
now be introduced in an analogous way by replacing $\enn(E;H_{\mathrm{cl}}^{(n)})$ by
$\enn(E;H^{(n)})$. For this to work we have to assume that the particle number $n$ is
so large that $\enn(E;H^{(n)})$ is three times differentiable. Below we will show
that at least for periodic potentials all derivatives up to order 3 are then
nonnegative.

Assume that for given $E>\inf\spec(H^{(n)})$ we have strict inequality
$d\enn(E;H^{(n)})/dE >0$. Define the probability measure
\begin{equation*}
d\nu_{E}(E^{\prime})=
\frac{1}{d\enn(E;H^{(n)})/dE}
\frac{d^2\enn(E^{\prime};H^{(n)})}{dE^{\prime^{\,^2}}}dE^{\prime}
\end{equation*}
on the interval $(\inf\spec(H^{(n)}),\,E]$. Note the close analogy with the classical
case \eqref{dnu1}. Continuing this analogy we introduce the function
\begin{equation*}
0\le\frac{1}{\cT(E^{\prime})}=
\frac{d}{dE^{\prime}}\ln
\frac{d^2}{dE^{\prime{\,^2}}}\enn(E^{\prime};H^{(n)})
\le\infty.
\end{equation*}
With $\langle\,\cdot\,\rangle^{\prime}_E$ denoting the resulting mean w.r.t.\
$d\nu_{E}$  we have $T(E)^{-1}=k\langle{\cT}^{-1}\rangle^{\prime}_E$. So again an
inverse temperature fluctuation is given by
\begin{equation*}
\Delta^{\prime}(T^{-1})(E)=
\left(k^2\langle{\cT}^{-2}\rangle^{\prime}_E
 -T(E)^{-2}\right)^{1/2}.
\end{equation*}
The relation \eqref{comp11} involving this fluctuation and the heat capacity again
carries over by simply replacing $\enn(E;H_{\mathrm{cl}}^{(n)})$ by
$\enn(E;H^{(n)})$.

As a first example we consider the free case $H^{(n)}_0$ (with or
without statistics) and for which the microcanonical entropy \eqref{sm} becomes
negative for small $E>0$ and $nd >2$ by \eqref{densn0}. As in the classical case by
\eqref{densn0} and
\eqref{T1} the resulting temperature is related to the energy by
\begin{equation}
\label{temp}
    E=\left(\frac{nd}{2}-1\right)kT(E).
\end{equation}
For $nd$ large this gives for the energy per particle
$E/n\approx d/2\,kT$ as should be expected. Again as in the classical case only
when $n=d=1$ the temperature so defined becomes negative. For $nd>6$ the inverse
temperature fluctuation is easily calculated to agree with the corresponding result
\eqref{deltafree} in the classical case.

The next relation gives the temperature difference between that of the theory with a
potential and that of the free theory at the same energy involving the potential only
through the scattering phase density,
\begin{equation}
\label{tdiff}
 T(E;H^{(n)})-T(E;H_0^{(n)})=\frac{1}{k}
\frac{\frac{d}{dE}\hat{\delta}(E;H^{(n)},H_0^{(n)})
-E\frac{d^2}{dE^2}\hat{\delta}(E;H^{(n)},H_0^{(n)})}
{c(nd)E^{nd/2-2}+
\frac{d^2}{dE^2}\hat{\delta}(E;H^{(n)},H_0^{(n)})}
\end{equation}
with $c(nd)=\Gamma(nd/2-1)^{-1}\pi(m/2\pi\hbar^2)^{nd/2}$. It is a consequence of
\eqref{densn0} and
\eqref{dens2}.
It would be interesting to analyze this expression and, in particular, the quantity
$\hat{\delta}(E;H^{(n)},H_0^{(n)})$ when $E$ approaches the bottom of the spectrum of
$H^{(n)}$.

Similarly to our discussion of the classical case we want to argue that at least for
periodic or random potentials, which in addition are twice differentiable, the heat
capacity is positive for all large energies. For this we use  the large energy
asymptotics of the phase shift \cite{S,Jensen,Robert} and on which we shall comment
below. In the present context this asymptotics takes the form (assuming in addition
that $V$ is twice differentiable in the periodic case or that the single site
potential $V_0$ is twice differentiable and has support strictly inside the unit cube
in the random case) we have
\begin{equation}
\label{deltan}
\frac{1}{|\Lambda|^n\pi}\delta(E;H_0^{(n)}+
V_{\Lambda^n}^{(n)},H_0^{(n)})
=\frac{1}{|\Lambda|^n}\enn(E;H^{(n)}_0)\left(a_{1}(n)E^{-1}+a_{2}(n)E^{-2}
+O(E^{-3})\right)
\end{equation}
with
\begin{equation*}
\begin{aligned}
a_{1}(n)&=-\frac{nd}{2}\int V_{\Lambda^n}^{(n)}(x)d^{nd}x = -
n\frac{nd}{2}\frac{1}{|\Lambda|}
\int V_{\Lambda}(x)d^{d}x, \\
a_{2}(n)&=\frac{nd}{4}\left(\frac{nd}{2}-1\right)
\int V_{\Lambda^n}^{(n)}(x)^2d^{nd}x\\
&=\frac{nd}{4}\left(\frac{nd}{2}-1\right)
\left[\frac{n}{|\Lambda|}\int
V_{\Lambda}(x)^2d^{d}x +\frac{n(n-1)}{|\Lambda|^2}\left(\int
V_{\Lambda}(x)d^{d}x\right)^2\right].
\end{aligned}
\end{equation*}
What has been proved so far is, e.g., that for periodic or random potentials in one
dimension there is an upper bound for the phase shift density of the form $const\cdot
E^{-1/2}$ \cite{KS3}. In three-dimensional case the three-term asymptotics of the
integrated density of states is known \cite{Karpeshina2}.

By the Schwarz inequality
\begin{equation}
\label{schwarz}
\frac{1}{|\Lambda|^2}\bigg(\int V_{\Lambda}(x)d^d x\bigg)^2
\le\frac{1}{|\Lambda|}\int V_{\Lambda}^2(x)d^d x
\end{equation}
with equality if and only if $V$ is constant on $\Lambda$. Both terms stay bounded
when $\Lambda\rightarrow\infty$. Let $\overline{V}^2$ and
$\overline{V^2}$, respectively, denote these limits as $\Lambda \rightarrow\infty$.
They agree with the quantities given by \eqref{Vav}. In the random case this limit is
deterministic, a trivial example of self-averaging.

Combined with \eqref{densn0} and \eqref{dens2}, and under the assumption that the
asymptotic expansion is preserved in the limit $\Lambda\rightarrow \infty$, relation
\eqref{deltan} gives the following asymptotic behavior of the integrated
density of states
\begin{eqnarray}
\label{deltacon1}
\enn(E;H_0^{(n)}+V^{(n)})&=&\enn(E;H^{(n)}_0)\bigg(1-E^{-1}n\frac{nd}{2}
\overline{V} \nonumber\\
&&+\frac{E^{-2}}{2}\frac{nd}{2}\left(\frac{nd}{2}-1\right)
\bigg[n\overline{V^2}
+n(n-1)\overline{V}^2\bigg]+O(E^{-3})\bigg).
\end{eqnarray}
Observe that $\hbar$ does not appear explicitly in this asymptotic expansion.
Therefore it makes sense to compare it with the corresponding classical case. Indeed,
when we multiply \eqref{deltacon1} by $h^{nd}$ the resulting derivative w.r.t.\ $E$
agrees with the derivative of $\enn(E;H_{\mathrm{cl}}^{(n)})$. This follows
\eqref{densn0} and \eqref{asymTV} and has the important consequence that to this
given order the asymptotic energy-temperature relation \eqref{asymTV1} as well as the
asymptotic relation \eqref{asymTV2} for the heat capacity and the conclusions thereof
carry over to this quantum case. Moreover, it is natural to conjecture the classical
limit relation
\begin{equation}
\label{classlim}
\lim_{\hbar\downarrow 0}\enn(E;H^{(n)})h^{nd}=\enn(E;H^{(n)}_{\mathrm{cl}}).
\end{equation}
For relations in this vein when the classical theory describes ergodic motion, see,
e.g., \cite{CdV,ST1,TU}. In addition, by the example in the previous section of a
classical model with negative specific heat, with an appropriate control of this
limit in \eqref{classlim} this could lead to a quantum model with negative specific
heat for small $\hbar$.

We turn to a discussion of the canonical ensemble in the quantum case. In analogy to
the classical case (see \eqref{ccan}) the Boltzmann-Gibbs canonical partition
function for $H^{(n)}_{\Lambda}$ may also be given in terms of
$N(E;H^{(n)}_{\Lambda})$ as
\begin{equation}
\label{qcan}
    Z(\beta;H^{(n)}_{\Lambda})=\tr_{L^2(\Lambda^n)}
(e^{-\beta H^{(n)}_{\Lambda}})=\int e^{-\beta
  E}dN(E;H^{(n)}_{\Lambda})
=\beta\int e^{-\beta
  E}N(E;H^{(n)}_{\Lambda})dE
\end{equation}
with mean energy
\begin{equation*}
 -\frac{d}{d\beta}\ln Z(\beta;H^{(n)}_{\Lambda})
=\frac{\int Ee^{-\beta E}dN(E;H^{(n)}_{\Lambda})}
{Z(\beta;H^{(n)}_{\Lambda})}.
\end{equation*}
Again by the Schwarz inequality it is a monotone increasing
function of the temperature.

So as in the classical case when $\Lambda\rightarrow\infty$ it therefore makes sense
to consider the following partition function per unit interaction volume
\begin{equation}
\label{qcan1}
 0\le z(\beta;H^{(n)})=\lim_{\Lambda\rightarrow\infty}
\frac{1}{|\Lambda|^n}Z(\beta;H^{(n)}_{\Lambda})
=\int e^{-\beta E}d\enn(E;H^{(n)})=\beta\int e^{-\beta E}\enn(E;H^{(n)})dE.
\end{equation}
The mean energy is then given as
\begin{eqnarray}
\label{cE1}
\bar{e}(\beta;H^{(n)})=-\frac{d}{d\beta}\ln z(\beta;H^{(n)})
&=&\frac{\int Ee^{-\beta E}d\enn(E;H^{(n)})} {z(\beta;H^{(n)})}\nonumber\\
&=&-\frac{1}{\beta}+\frac{\beta}{z(\beta;H^{(n)})}
\int Ee^{-\beta E}\enn(E;H^{(n)})dE,
\end{eqnarray}
which is not necessarily positive since $\enn(E;H^{(n)})$ may be non-zero for $E<0$.
Again by the Schwarz inequality the mean energy for the canonical
distribution is a monotone increasing function of the temperature. For the free
case we obtain the familiar relation
\begin{equation}
\label{qcan0}
    z(\beta;H^{(n)}_0)=\left(\frac{m}{2\pi\hbar^2\beta}\right)^{nd/2}
=\frac{1}{h^n}(2\pi mkT)^{nd/2}
\end{equation}
and hence the energy-temperature relation $\bar{e}(\beta;H^{(n)}_0)=(nd/2)kT$ which
agrees with the one in the microcanonical ensemble \eqref{temp} when $n$ is large. By
\eqref{dens2} (compare also with the relation \eqref{Krein})
in the one-particle case, we obtain the relation
\begin{equation}
\label{Krein1}
z(\beta;H^{(n)})-z(\beta;H_0^{(n)})=\frac{1}{\pi}\int e^{-\beta
E}d\hat{\delta}(E;H^{(n)},H_0^{(n)})
=\frac{\beta}{\pi}\int
e^{-\beta E}\hat{\delta}(E;H^{(n)},H_0^{(n)})dE.
\end{equation}
The relations \eqref{qcan0} and \eqref{Krein1} give
\begin{eqnarray}\label{ediff}
\lefteqn{\bar{e}(\beta;H^{(n)})-\bar{e}(\beta;H^{(n)}_0)=-\frac{d}{d\beta}
\ln \frac{z(\beta;H^{(n)})}{z(\beta;H^{(n)}_0)}}\nonumber\\
&=&-\frac{d}{d\beta}
\ln \bigg(1+\left(\frac{2\pi\hbar^2\beta}{m}\right)^{nd/2}\frac{1}{\pi}
\int e^{-\beta E}d\hat{\delta}(E;H^{(n)},H^{(n)}_0)\bigg)\nonumber\\
&=&-\bigg(\frac{2\pi\hbar^2\beta}{m}\bigg)^{nd/2}
\bigg(1+\left(\frac{2\pi\hbar^2\beta}{m}\right)^{nd/2}
\frac{1}{\pi}
\int e^{-\beta E}d\hat{\delta}(E;H^{(n)},H^{(n)}_0)\bigg)^{-1}\nonumber\\
&&\cdot\frac{1}{\pi}\int\left(\frac{nd}{2}\beta^{-1}-E\right) e^{-\beta
E}d\hat{\delta}(E;H^{(n)},H^{(n)}_0).
\end{eqnarray}
This relation corresponds to the temperature difference \eqref{tdiff} in the
microcanonical description and again the right hand side involves the potential only
through the scattering phase density. We do not know at present
under which conditions one may take the limit $T\rightarrow 0$ in \eqref{ediff} and
what the result should be. As in the microcanonical case this presumably requires an
analysis of the scattering phase density near the bottom of the spectrum of $H^{(n)}$
which equals $n \min(\inf \spec (H),0)$ (see the remark after \eqref{sas} below).

We note in this context that a theorem of Fumi \cite{Fu} (see also \cite{Br,FN,
deWitt,M}) relates the shift of the ground state energy due to a single impurity of a
system of non-interacting fermions to the scattering phase caused by this impurity by
zero temperature. We may recover Fumi's theorem now in the form
of a statement on the shift of the mean energy density for non-interacting electrons,
moving in a periodic or random potential $V$, from the general relation
\eqref{dens1} in the following way. For given $V$ and given electron
density $\ennn$ let the Fermi energy $E_F=E_F(\ennn;H)$ be given as the solution to
the equation $\ennn/2=\enn(E_F;H)$, where the factor $1/2$ accounts for the spin.
Thus $E_F(\ennn;H_0) = (\ennn h^d
\Gamma(d/2+1)/2)^{2/d}/(2\pi m)$ for the free theory. Then the ground state energy
density $\Omega_F$ is given as
\begin{equation}
\label{omega}
\Omega_F(\ennn;H)=2\int_{\inf \spec (H)}^{E_F(\ennn;H)}Ed\enn(E;H)
=-2\int_{\inf \spec (H)}^{E_F(\ennn;H)}\enn(E;H)dE+E_F(\ennn;H)\ennn.
\end{equation}
Here again the factor 2 accounts for the spin and the last relation follows by
partial integration. For the free theory we have
$\Omega_F(\ennn;H_0)=d/(2(d/2+1))E_F(\ennn;H_0)\ennn$.

We recall that $\enn(E;H)$ is a monotone increasing function in $E$ but
for periodic $V$ it is constant outside the spectral bands, so $E_F$ may not be
uniquely defined if $\ennn$ is such that $E_F(\ennn;H)$ lies outside the bands. But
by the same observation and the first equality in \eqref{omega} $\Omega_F(\ennn;H)$
is still unique. For random potentials with no gaps in the spectrum the deterministic
quantity $\enn(E;H)$ is strictly increasing, so there is no problem then with the
definition of $E_F(\ennn;H)$.

{}From \eqref{omega} we obtain the following relation for the shift of the mean
energy density w.r.t.\ the free theory as
\begin{eqnarray*}
\Omega_F(\ennn;H)-\Omega_F(\ennn;H_0)
&=&2\int_{\inf \spec (H)}^{E_F(\ennn;H)}Ed\enn(E;H)-
2\int_{0}^{E_F(\ennn;H_0)}Ed\enn(E;H_0)\nonumber\\ &=&-2\int_{\inf \spec
(H)}^{E_F(\ennn;H)}\enn(E;H)dE- 2\int_{0}^{E_F(\ennn;H_0)}\enn(E;H_0)dE\nonumber\\
&&+(E_F(\ennn;H)-E_F(\ennn;H_0))\ennn\nonumber\\ &\cong&-\frac{2}{\pi}\int_{\min
(\inf \spec (H),0)}^{E_F(\ennn;H_0)}
\hat{\delta}(E;H,H_0)dE
\end{eqnarray*}
in the approximation where $E_F(\ennn;H)\cong E_F(\ennn;H_0)$, which should be valid
for weak potentials. Again we note that as in the case of the extension of Friedel's
theorem the contribution by several impurities is not just the sum of the
contributions of the single impurities. The correct analysis of the contributions by
random impurities was performed in our work \cite{KS0,KS1}.

Returning to our general discussion of the canonical ensemble an entropy may be
defined in the following way. With $\rho(\beta;H^{(n)}_{\Lambda})=\exp(-\beta
H^{(n)}_{\Lambda})/Z(\beta;H^{(n)}_{\Lambda})$ being the density matrix (i.e., $0\le
\rho(\beta;H^{(n)}_{\Lambda})$ and
$\tr_{L^2(\Lambda^n)}\rho(\beta;H^{(n)}_{\Lambda})=1$) associated to the canonical
theory for $H^{(n)}_{\Lambda}$ at temperature $T$, let
\begin{eqnarray}
\label{vNent}
0\le S(\beta;H^{(n)}_{\Lambda})&=&-k\,\tr_{L^2(\Lambda^n)}
\left(\rho(\beta;H^{(n)}_{\Lambda})\ln \rho(\beta;H^{(n)}_{\Lambda})\right)
\nonumber\\
&=&k\ln Z(\beta;H^{(n)}_{\Lambda})+\frac{k\beta}{Z(\beta;H^{(n)}_{\Lambda})}
\int Ee^{-\beta E}dN(E;H^{(n)}_{\Lambda})\nonumber\\
&=&-k\int Z(\beta;H^{(n)}_{\Lambda})^{-1}e^{-\beta E}
\ln\left(Z(\beta;H^{(n)}_{\Lambda})^{-1}e^{-\beta E}\right)
dN(E;H^{(n)}_{\Lambda})\nonumber\\ &=&-k\beta^2\frac{d}{d\beta}
\left(\frac{1}{\beta}\ln Z(\beta;H^{(n)}_{\Lambda})\right)
\ =\ k\beta^2\frac{d}{d\beta}F(\beta;H^{(n)}_{\Lambda})
\end{eqnarray}
be the resulting von Neumann entropy \cite{N}. Also
\begin{equation*}
F(\beta;H^{(n)}_{\Lambda})=-\frac{1}{\beta}\ln Z(\beta;H^{(n)}_{\Lambda})
\end{equation*}
is the Helmholtz free energy. Recall that $S(\beta;H^{(n)}_{\Lambda})\ge 0$ holds
because $-x\ln x\ge 0$ for $0\le x\le 1$ and all eigenvalues of
$\rho(\beta;H^{(n)}_{\Lambda})$ lie between 0 and 1. By the second relation in
\eqref{vNent} the limit
\begin{equation}
\label{sc}
s(\beta;H^{(n)})=\lim_{\Lambda\rightarrow\infty}
\left(S(\beta;H^{(n)}_{\Lambda})-kn\ln |\Lambda|\right)
\end{equation}
exists and is given as
\begin{eqnarray}
\label{vNent1}
s(\beta;H^{(n)})&=&k\ln z(\beta;H^{(n)})+\frac{k\beta}{z(\beta;H^{(n)})}
\int Ee^{-\beta E}d\enn(E;H^{(n)})\nonumber\\
&=&-k\int z(\beta;H^{(n)})^{-1}e^{-\beta E}
\ln\left(z(\beta;H^{(n)})^{-1}e^{-\beta E}\right)d\enn(E;H^{(n)})\nonumber\\
&=&-k\beta^2\frac{d}{d\beta}\left(\frac{1}{\beta}\ln z(\beta;H^{(n)})\right).
\end{eqnarray}
We interpret this as the entropy per unit interaction volume. Since
$S(\beta;H^{(n)}_{\Lambda})$ is a monotone decreasing function in $\beta$ so is
$s(\beta;H^{(n)})$ by \eqref{sc}.

By \eqref{qcan0} for the free case we have
\begin{equation*}
s(\beta;H^{(n)}_0)=\frac{nd}{2}k\left(1+\ln\frac{m}{\hbar^2\beta}\right)
\end{equation*}
so in particular $s(\beta;H^{(n)}_0)$ tends to $+\infty$ as $\beta\rightarrow 0$ and
to $-\infty$ as $\beta\rightarrow \infty$. In analogy to the microcanonical case
\eqref{sm} the last property is again an undesired feature, which it shares with the
classical theory (see, e.g., \cite{W}). The first property continues to hold in the
presence of a potential $V$, which in the present context we will assume to be
bounded. The way to see this is to consider the difference
$s(\beta;H^{(n)})-s(\beta;H^{(n)}_0)$. It has an asymptotic expansion in $\beta$ for
$\beta$ small, which is obtained as follows. Using the so called heat kernel
expansion one can show (see, e.g., \cite{Gi} and the references given there) that the
following asymptotic expansion is valid
\begin{eqnarray}
\label{heat}
\lefteqn{\frac{1}{|\Lambda|^n}Z(\beta;H^{(n)}_{\Lambda})
=\left(\frac{m}{2\pi\hbar^2\beta}\right)^{nd/2}
\Bigg(1-n\beta \frac{1}{|\Lambda|}\int V_{\Lambda}(x)d^d x}\nonumber\\
&&\qquad\qquad\quad + \frac{\beta^2}{2} \left[n \frac{1}{|\Lambda|}
\int V_{\Lambda}^2(x)d^d x
+ \frac{n(n-1)}{|\Lambda|^2}\left(\int V_{\Lambda}(x)d^dx\right)^2\right]
+O(\beta^3)\Bigg).
\end{eqnarray}
Here we have assumed that $\Lambda$ is a finite union of unit cubes. In addition in
the periodic case the potential $V$ is assumed to be twice differentiable and in the
random case \eqref{stopo} the single site potential $V_0$ with support in the unit
cube centered at the origin is also twice differentiable and vanishes  near the
boundary. In case $V$ and $V_0$ have higher order derivatives the asymptotic
expansion \eqref{heat} also extends to higher orders. Note that apart from the
pre-factor Planck's constant does not appear. This is a consequence of the
assumptions just made on $V$, $V_0$ and on $\Lambda$. In general,
powers of $\hbar$ appear combined with derivatives of $V$ and
$V_{\omega}$ of the same order within integrals over $\Lambda$. Integration over
$\Lambda$ therefore gives boundary contributions which vanish.

Assuming again that this asymptotic expansion is preserved in the limit
$\Lambda\rightarrow\infty$, we obtain
\begin{equation}
\label{heat1}
z(\beta;H^{(n)})
=z(\beta;H^{(n)}_0)
\bigg(1-n\beta\overline{V}
+ \frac{\beta^2}{2} \left[n \overline{V^2}
+n(n-1)\overline{V}^2\right]+O(\beta^3)\bigg).
\end{equation}
Again up to a factor $h^{nd}$ this agrees with \eqref{asymeef}. Using \eqref{qcan1}
we see that \eqref{heat1} is compatible with \eqref{deltacon1}, i.e.,  the
Laplace-Stieltjes transform of \eqref{deltacon1} gives \eqref{heat1}. As a matter of
fact, the conjecture made in \cite{S} on the asymptotic behavior of the phase shift
(which in the present context is \eqref{deltan}) was just made to fit in this way
with the ``heat kernel'' expansion at large temperatures (small $\beta$) of
\begin{equation*}
\frac{1}{|\Lambda|^n}\tr_{L^2(\R^{nd})}
\left(e^{-\beta(H^{(n)}_0+V^{(n)}_{\Lambda^n})}
-e^{-\beta H^{(n)}_0}\right).
\end{equation*}
For comparison note that in general no asymptotic expansion beyond the famous Weyl
term for the number of eigenvalues of an elliptic differential operator on a compact
manifold like $N(E;H^{(n)}_{\Lambda})$ is known. The best result in this direction is
due to H\"{o}rmander \cite{Hoe}. Although related, the conjecture \eqref{deltan} is of a
different nature since instead of the operator $H^{(n)}_{\Lambda}$ it involves the
operator $H^{(n)}_0+V^{(n)}_{\Lambda^n}$.

Observe that apart from the overall factor so far $\hbar$ does not appear in the
asymptotic expansion of $z(\beta;H^{(n)})$ to this order. This is of course in
agreement with our corresponding result in the microcanonical setup. Although we
shall not need it, we remark that for finite range, smooth potentials $W$ in $d$
dimensions the trace of the heat kernel $\exp\{-\beta(H_0+W)\}$ multiplied by $h^d$
and for fixed $\beta$ has a well known asymptotic expansion in (positive) powers of
$h$ (see, e.g., \cite{LL}) which has been shown to be a complete asymptotic expansion
in \cite{HeRo,ST}. For the asymptotics of the integrated density of states for
periodic Schr\"{o}dinger operators we refer to \cite{Karpeshina2}.

Relation \eqref{heat1} results in the following relation for the entropy
\begin{equation}\label{sas}
s(\beta;H^{(n)})-s(\beta;H^{(n)}_0) = -k\beta^2\frac{d}{d\beta}\frac{1}{\beta}
\ln \frac{z(\beta;H^{(n)})}{z(\beta;H^{(n)}_0)}
=-kn\frac{\beta^2}{2}\left(\overline{V^2}
-\overline{V}^2\right) + O(\beta^3).
\end{equation}
By \eqref{schwarz} the first term on the right hand side of \eqref{sas} is strictly
negative if $V\neq 0$. We do not now about a similar asymptotic
expansion in $1/\beta$, which is uniform in $\Lambda$. The problem arises because as
mentioned above it is difficult to establish the asymptotic behavior of \eqref{ediff}
for small $T$. Note that for fixed $\Lambda$ one has, e.g.,
\begin{equation*}
\ln Z(\beta;H^{(n)}_{\Lambda})=-\beta \epsilon_0+\ln \dim P_0
+\frac{\dim P_1}{\dim P_0}e^{-\beta(\epsilon_1-\epsilon_0)}
+O(e^{-\beta(\epsilon_2-\epsilon_1)}),
\end{equation*}
where $\epsilon_0<\epsilon_1<\epsilon_2<...$ are the different eigenvalues of
$H^{(n)}_{\Lambda}$ and $P_i$ the orthogonal projections onto the corresponding
eigenspaces such that $\dim P_i$ are the multiplicities of the eigenvalues. For fixed
$i$ the difference $\epsilon_i-\epsilon_0$ tends to zero at
least as $|\Lambda|^{1/d}$ as $\Lambda\rightarrow\infty$.

Define
\begin{equation*}
f(\beta;H^{(n)})
=-\frac{1}{\beta}\lim_{\Lambda\rightarrow\infty}
(F(\beta;H^{(n)}_{\Lambda})-n|\Lambda|)
=-\frac{1}{\beta}\ln z(\beta;H^{(n)})
\end{equation*}
to be the Helmholtz free energy per interaction volume. {}From \eqref{cE1} and
\eqref{vNent1} we obtain the relation between the mean energy, the entropy and the
free energy, familiar in the usual context of the canonical ensemble (compare with
the same relation \eqref{meef1} in the classical case)
\begin{equation}
\label{cef}
\bar{e}(\beta;H^{(n)})=Ts(\beta;H^{(n)})+f(\beta;H^{(n)}).
\end{equation}
By \eqref{heat} we have
\begin{equation*}
f(\beta;H^{(n)})-f(\beta;H^{(n)}_0)=n\overline{V}-n\frac{\beta}{2}
\left[\overline{V^2}
-\frac{1}{2}\overline{V}^2\right]+O(\beta^2)
\end{equation*}
for small $\beta$ and where $f(\beta;H^{(n)}_0)$ may be read off \eqref{qcan0}. The
mean energy has the asymptotic form
\begin{equation*}
\bar{e}(\beta;H^{(n)})=\frac{nd}{2}\beta^{-1}+n\overline{V}
-\frac{n\beta}{2} \left[\overline{V^2}
-\overline{V}^2\right]+O(\beta^2)
\end{equation*}
which agrees with the asymptotic behavior \eqref{mteasy} in the classical canonical
description. Therefore the inverse asymptotic relation \eqref{tmeasy} carries over
and so for large $E$ and $n$ these microcanonical and canonical descriptions are
approximately equal. In other words in analogy to \eqref{clinv} we have the
approximative relations which are inverse to each other
\begin{equation*}
T\approx T(\bar{e}(\beta=1/kT;H^{(n)})),\qquad E\approx
\bar{e}(\beta=1/kT(E;H^{(n)});H^{(n)}).
\end{equation*}

Also in this canonical setup the heat capacity is defined as
\begin{equation}
\label{cv}
c_{\mathrm{v}}(\beta;H^{(n)})=\frac{d}{dT}\bar{e}(\beta;H^{(n)})=
\frac{1}{kT^2}\Delta\bar{e}(\beta;H^{(n)})^2,
\end{equation}
which is $\geq 0$ by the discussion above and which equals $knd/2$ for all $\beta$ in
the free case as it should. As in \eqref{varebar}
\begin{equation}
\label{varebar1}
\Delta\bar{e}(\beta;H^{(n)})=
\left(\frac{d^2}{d\beta^2}\ln z(\beta;H^{(n)})\right)^{1/2}
\end{equation}
is the energy fluctuation. So for small $\beta$ we obtain
\begin{equation}
\label{cvas}
c_{\mathrm{v}}(\beta;H^{(n)})=\frac{knd}{2}+\frac{kn\beta^2}{2}\left[\overline{V^2}
-\overline{V}^2\right]+O(\beta^3)
\end{equation}
and thus for small $\beta$ the heat capacity increases when an external potential $V$
is switched on. By \eqref{cv} relation \eqref{cvas} in turn gives the following
asymptotic behavior for the energy fluctuation
\begin{equation*}
\Delta\bar{e}(\beta;H^{(n)})=
\sqrt{\frac{nd}{2}}\frac{1}{\beta}\left(1+\frac{\beta^2}{d}\left[\overline{V^2}
-\overline{V}^2\right]+O(\beta^3)\right),
\end{equation*}
which by \eqref{varebar1} may also be obtained from \eqref{heat1} and which agrees
with its classical counterpart \eqref{ebarascl}.

In analogy to our analysis in the classical case (see the discussion following
\eqref{cg}) and by the same arguments (involving the assumption that the heat
capacity is sufficiently large and positive), also in this quantum context the
canonical and the microcanonical ensemble both give the same energy-temperature
relation and the same heat capacities. In particular the variance $\alpha$ of the
approximating Gauss distribution is just the square of the energy fluctuation
\eqref{varebar1}, which in turn is related to the heat capacity by \eqref{cv}.
Indeed, by observing the necessary differentiability conditions the discussion can be
carried over \emph{verbatim}.

We now want to establish that for given one particle theory with a periodic external
potential smoothness of $\enn(E;H^{(n)})$ increases with the particle number $n$ and
that the derivatives are all nonnegative. For this the next relation will become
relevant. Observe first that for all $n\ge 2,1 \le k\le n-1,\,\Lambda$ and $E$
\begin{equation*}
N(E;H^{(n)}_{\Lambda})=\int
N(E-E^{\prime};H_{\Lambda}^{(k)})dN(E^{\prime};H_{\Lambda}^{(n-k)}).
\end{equation*}
This follows easily from the identity
\begin{equation*}
\Theta(E-\epsilon-\epsilon^{\prime})
=\int\Theta(E-E^{\prime}-\epsilon)\delta(E^{\prime}
-\epsilon^{\prime})dE^{\prime}
=\int\Theta(E-E^{\prime}-\epsilon)d\Theta(E^{\prime}
-\epsilon^{\prime})
\end{equation*}
and from $H_{\Lambda}^{(n)}=
H_{\Lambda}^{(k)}\otimes\1_{L^2(\Lambda^{n-k})}+\1_{L^2(\Lambda^k)}\otimes
H_{\Lambda}^{(n-k)}$ and $\tr_{L^2(\Lambda^n)}(A\otimes B)=\tr_{L^2(\Lambda^k)}(A)
\tr_{L^2(\Lambda^{n-k})}(B)$. Therefore the integrated density of states for the
$n$-particle theory obeys a similar relation, i.e., it may be written as  a
Riemann-Stieltjes convolution
\begin{equation}
\label{conv2}
\enn(E;H^{(n)})=\int
\enn(E-E^{\prime};H^{(k)})d\enn(E^{\prime};H^{(n-k)})=\enn(\cdot;H^{(k)})*_S
\enn(\cdot;H^{(n-k)})(E).
\end{equation}
In particular $\enn(E;H^{(n)})$ is an $(n-1)$-fold iterated Riemann-Stieltjes
convolution of the one-particle integrated density of states $\enn(E;H^{(n=1)})$ with
itself. Here we used the assumption that $V\ge -c$ for some $c\ge 0$, such that
$H^{(n)}\ge -nc$ and hence $\enn(E,H^{(n)})=0$ for all $E<-nc$. Let
$E^{(n)}_{\mathrm{thresh}}>-cn$ be the infimum of all $E$ such that
$\enn(E;H^{(n)})\neq 0$. Since $\enn(E;H)$ is H\"{o}lder continuous and nonnegative, the
Riemann-Stieltjes convolution of $\enn(E;H)$ with itself exists and is also H\"{o}lder
continuous. This is easily established with techniques used in, e.g., \cite{G,DN}.
Relation \eqref{conv2} implies
\begin{equation*}
z(\beta;H^{(n)})= z(\beta;H^{(k)})z(\beta;H^{(n-k)})
\end{equation*}
giving in particular $z(\beta;H^{(n)})=z(\beta;H)^n$ and the obvious relation
$f(\beta;H^{(n)})=n f(\beta; H)$ for the free energy. Relations of this type are of
course familiar in the ordinary quantum canonical ensemble formulation of
noninteracting particles. Thus we have, e.g., for the partition function given by
\eqref{qcan}
\begin{equation*}
Z(\beta;H^{(n)}_{\Lambda})= Z(\beta;H^{(n-k)}_{\Lambda})Z(\beta;H^{(k)}_{\Lambda}).
\end{equation*}
With reasonable assumptions on $\enn(E;H)$ concerning the van Hove singularities when
$V$ is a periodic potential and which hold in $d=1$ dimension and for the Kronig-
Penney model (see, e.g., \cite{A}, for a survey of spectral properties for periodic
potentials in general see \cite{Karpeshina}) we will show in the appendix that
$\enn(E;H^{(n=2)})$ is actually continuously differentiable except for a discrete set
of energies (without finite accumulation points) where, however, the right and left
derivatives exist. As a consequence $\enn(E;H^{(n=3)})$ is continuously
differentiable at all energies. Using \eqref{conv2} by complete induction we obtain
that $\enn(E;H^{(n)})$ is $[(n-1)/2]$ times continuously differentiable in $E$ with
the representation for the $l$'th derivative $(l\le [(n-1)/2])$ as
\begin{eqnarray}
\label{conv4}
\enn^{(l)}(E;H^{(n)})&=&\int
\enn^{(l-1)}(E-E^{\prime};H^{(n-2)})d\enn^{\prime}(E^{\prime};H^{(2)})\nonumber\\
&=&\enn^{(l-1)}(\cdot;H^{(n-2)})*_S \enn^{\prime}(\cdot;H^{(2)})(E).
\end{eqnarray}
Also from \eqref{conv4} it follows again by complete induction in $l$ that these
derivatives are all nonnegative. In particular $\enn(E;H^{(n)})$ is concave in $E$
for $n\ge 5$. This increase of smoothness with $n$ is related to the known increase
of smoothness with the dimension $d$ of space. Indeed write $H_0^d$ to denote the
dependence of the free Hamiltonian on the space dimension. Then for given external
one particle potential $V$ we have $(H_0^d+V)^{(n)}=H_0^{nd}+V^{(n)}$, where as in
the classical case the potential $V^{(n)}$ is given as $V^{(n)}(x)=\sum_i V(x_i)$ for
$x=(x_1,x_2,\ldots,x_n)\in \R^{nd}$ and with $x_i\in \R^d$. Note that periodicity
extends in the sense that e.g.\ $V^{(n)}(x+j)=V^{(n)}(x)$ holds for any $j\in
\Z^{nd}$ and $x \in \R^{nd}$, whenever $V(y+i)=V(y)$ for all $i\in
\Z^{d}$ and $y \in \R^{d}$.

To sum up by our preceding discussion for one particle Schr\"{o}dinger operators with
periodic potentials we therefore may define a temperature for the microcanonical
ensemble for all $H^{(n)},\,n \ge 5$. Since the third derivative of
$\enn(E;H^{(n)}),n\ge 7$ is non-negative it can not vanish identically on any
interval [$E^{(n)}_{\mathrm{thresh}},E$]. To see this, assume the contrary. Then by
integrating 3 times we see that $\enn(E;H^{(n)})$ also would vanish identically on
this interval, which is not possible. So again by integrating, the first and second
derivatives of $\enn(E;H^{(n)}),n\ge 5$ are strictly positive for any
$E>E^{(n)}_{\mathrm{thresh}}$. So we conclude that the microcanonical temperature for
$H^{(n)},n\ge 5$ is actually positive, finite and a continuous function of $E$.
Similarly, since the fourth derivative of the density of states for $H^{(n)},n\ge 9$
is non-negative, the second derivative is concave in $E$. However, this is not
sufficient to conclude that $1/T$ is a concave function of $E$ whenever $n\ge 9$ and
presently we do not know sufficient conditions on the potential $V$ which ensure
this.

We expect similar results to be valid for the case of stochastic potentials, where it
is known that the integrated density of states is Lipschitz continuous, i.e., H\"{o}lder
continuous of index $1$ (see \cite{CH}), so one has smoother properties than in the
periodic case. For easier, recent proofs that they are H\"{o}lder continuous of any index
smaller than $1$, see \cite{CHN,KS2}. Unfortunately, in general the Riemann-Stieltjes
convolution does not improve H\"{o}lder continuity or any other similar kind of
regularity (see, e.g., \cite{G,DN}). It would be interesting to find additional
properties of the integrated density of states which allows one to conclude that
their Riemann-Stieltjes convolution improves regularity.

So far we have not taken Bose-Einstein or Fermi-Dirac statistics into account. To do
this observe first that the permutation group $S_n$ of $n$ elements has a canonical
unitary representation $\pi\rightarrow U(\pi),\,\pi\in S_n$ on $\otimes_n L^2(\R^d)$.
A similar observation is valid when $\R^d$ is replaced by $\Lambda$. The orthogonal
projections onto the subspaces of symmetric (+) and antisymmetric (-) wave-functions
$\otimes_{n,\pm}L^2(\R^d)$ are given as
\begin{equation}
\label{stat}
P_{\pm}^{(n)}=\frac{1}{n!}\sum_{\pi\in S_n}(sign\;\pi)^{(1\mp 1)/2}U(\pi)
\end{equation}
respectively, i.e., $\otimes_{n,\pm}L^2(\R^d)=P_{\pm}^{(n)}\otimes_{n}L^2(\R^d)$.
This gives
$\tr_{\otimes_{n,\pm}L^2(\R^d)}(A)=tr_{\otimes_{n}L^2(\R^d)}(P_{\pm}^{(n)}A)$ and
similarly with $\R^d$ being replaced by $\Lambda$. Then one may show that the
resulting density of states are given by
\begin{equation}
\label{stat1}
n_{\pm}(E;H^{(n)})=\lim_{\Lambda\rightarrow\infty}
\frac{1}{|\Lambda|^n}\tr_{L^2(\Lambda^n)}
\left(P_{\pm}^{(n)}\Theta(E-H_{\Lambda}^{(n)})\right)=\frac{1}{n!}
\enn(E;H^{(n)}),
\end{equation}
i.e., when inserting \eqref{stat} into the middle term in \eqref{stat1} the
contributions from $\pi\neq id$ vanish in the limit. In the periodic case and in
analogy to \eqref{densn1} one also has
\begin{equation*}
n_{\pm}(E;H^{(n)})=\tr_{L^2(\R^{nd})}
\left(\chi_{\Lambda_0^n}P_{\pm}^{(n)}\Theta(E-H^{(n)})\right).
\end{equation*}
In particular the resulting temperature in the microcanonical ensemble does not
depend on the statistics. Similarly the partition function in the canonical ensemble
is modified by a factor $1/n!$.

To conclude this section we briefly describe the associated grand canonical ensemble
for both statistics using the integrated density of states. In a finite volume
$\Lambda$ the partition function for the grand canonical ensemble is given as
\begin{equation*}
Z_{\pm}(\beta,\mu;H)=1+\sum_{n=1}^{\infty}e^{-\mu\beta n}\tr_{L^2(\Lambda^n)}
\left(P_{\pm}^{(n)}e^{-\beta H^{(n)}}\right).
\end{equation*}
with $\mu$ being the chemical potential. Standard arguments therefore give
\begin{eqnarray*}
\ln Z_{\pm}(\beta,\mu;H)&=&\tr_{L^2(\Lambda^n)}\left(1\mp
e^{-\beta(\mu+H_{\Lambda}^{(n=1)})}\right)^{\mp 1}=\int \left(1\mp
e^{-\beta(\mu+E)}\right)^{\mp 1}dN(E;H_{\Lambda}^{(n=1)}).\nonumber\\ &&
\end{eqnarray*}
Hence $z_{\pm}(\beta,\mu;H)$ may be defined in terms of the one-particle integrated
density of states by
\begin{equation*}
\ln z_{\pm}(\beta,\mu;H)=\lim_{\Lambda\rightarrow\infty}
\frac{1}{|\Lambda|}\ln Z_{\pm}(\beta,\mu;H)=\int \left(1\mp
e^{-\beta(\mu+E}\right)^{\mp 1}d\enn(E;H^{(n=1)}).
\end{equation*}
The usual relations for the mean energy and the mean particle number may now be
obtained easily.

\begin{appendix}
\section{}
\renewcommand{\theequation}{\mbox{\Alph{section}.\arabic{equation}}}
\setcounter{equation}{0}

In the the first part of the appendix we will establish increasing regularity for the
quantum integrated density of states with increasing particle number in the case of a
periodic potential. In the second and last part we will show in an example how
randomness may serve to increase the smoothness of the classical integrated density
of states already in the one particle case.

For periodic potentials $V$, where bands appear, it is well known that the
one-particle integrated density of states $\enn(E;H^{(n=1)})$ is constant for $E$
inside the gaps, smooth inside the bands and behaves like $|E-E_{i,\pm}|^{1/2}$ at an
upper $E_{i,+}$ or lower edge $E_{i,-}$ of a band, so in particular it is H\"{o}lder
continuous of index $1/2$ everywhere. More precisely we make the following
assumption, which has been proven rigorously in $d=1$ dimension, see
\cite{Magnus:Winkler}. Write the bands (finitely or infinitely many) as the closed,
pairwise disjoint intervals $[E_{i,-},E_{i,+}]$. There are smooth functions
$n_{i,\pm}(E;H),\; E\ge 0$ small, with $n_{i,\pm}(E=0;H)=0$ such that for $E$ near
$E_{i,\pm}$
\begin{equation}
\label{ass}
\enn(E;H)=\left\{\begin{array}{ccc} |E-E_{i,\pm}|^{1/2}
n_{i,\pm}(\mp(E-E_{i,\pm});H)+c_{i,\pm} &\mbox{if}&\mp(E-E_{i,\pm})\ge 0\\
  c_{i,\pm}&\mbox{if}&\pm(E-E_{i,\pm})\ge 0
\end{array}\right.,
\end{equation}
where $c_{i,\pm}=\enn(E_{i,\pm};H)$. In particular $\enn(E;H)$ is H\"{o}lder continuous
of index $1/2$. Also away from the ends $E_{i,\pm}$ of the bands the function
$\enn(E;H)$ is supposed to be smooth. We note that for $d>1$ the Bethe-Sommerfeld
conjecture is actually true: As established by Skriganov and Karpeshina, there are
only finitely many gaps (see, e.g., \cite{Ku} for an extensive list of references and
\cite{Karpeshina}).

With these assumptions we will show that $\enn(E;H^{(n=2)})$ is continuously
differentiable in $E$ from the right and from the left. Also the right and left
derivatives agree except for the discrete set of energies of the form
$E=E_{i,\pm}+E_{j,\pm}$. In other words the derivative of $\enn(E;H^{(n=2)})$ is
possibly discontinuous at these energies. Furthermore we will show that
$\enn(E;H^{(n=3)})$ is differentiable in $E$ for all $E$ with a derivative which is
H\"{o}lder continuous of index $1/2$.

For the proof let $0\le \chi_{i,\pm}(E)\le 1$ be smooth functions which are equal to
one near $E_{i,\pm}$ and where \eqref{ass} holds. They may be chosen to have
non-overlapping support, i.e., any product of two different $\chi_{i,\pm}(E)$'s is
zero. Set $0\le\chi(E)=\sum_{i,\pm}\chi_{i,\pm}(E)\le 1$ and write
$\enn_1(E;H)=(1-\chi(E))\enn(E;H),\enn_2(E;H)=\chi(E)\enn(E;H)$ such that
$\enn(E;H)=\enn_1(E;H)+\enn_2(E;H)$. Then $\enn_1(E;H)$ is smooth everywhere and
$\enn_2(E;H)$ vanishes outside small intervals around the points $E_{i,\pm}$. With
this decomposition of $\enn(E;H)$ we obtain
\begin{equation}
\label{decomp}
\enn(E;H^2)=\enn_1(\cdot;H)*_S\enn_1(\cdot;H)(E)+2\enn_1(\cdot;H)
*_S\enn_2(\cdot;H)(E)
+\enn_2(\cdot;H)*_S\enn_2(\cdot;H)(E).
\end{equation}

The first two terms in \eqref{decomp} are easily seen to be continuously
differentiable w.r.t.\ $E$, so we only have to show that the third term is
continuously differentiable. It suffices to consider any term of the form
\begin{equation}
\label{int1}
\int\chi_{i,\pm}(E-E^{\prime})\enn(E-E^{\prime};H)
d(\chi_{j,\pm}(E^{\prime})\enn(E^{\prime};H)),
\end{equation}
since for given $E$ there are only finitely many $E_{i,\pm}$'s with $E_{i,\pm}\le E$.
Now we use the assumption \eqref{ass} and obtain
\begin{eqnarray*}
d(\chi_{l,\pm}(E)\enn(E;H))
&=&\Theta(\mp(E-E_{l,\pm}))\left(|E-E_{l,\pm}|^{1/2}f_{l,\pm,1}(E)+
|E-E_{l,\pm}|^{-1/2}f_{l,\pm,2}(E)\right)dE \\ &&+f_{l,\pm,3}(E)dE
\end{eqnarray*}
$l=i$ or $j$, where the $f_{l,\pm,k}$'s, $k=1,2,3$ are smooth with support contained
in the support of $\chi_{l,\pm}$. We insert this into the formal derivative of
\eqref{int1} and obtain 9 terms all of which should be finite integrals. The
potentially most dangerous one and on which we shall concentrate is
\begin{equation}
\label{int2}
\int\Theta(\mp(E-E^{\prime}-E_{i,\pm}))\Theta(\mp(E^{\prime}-E_{j,\pm}))
|E-E^{\prime}-E_{i,\pm}|^{-1/2}|E^{\prime}-E_{j,\pm}|^{-1/2}
f_{i,\pm,2}(E-E^{\prime})f_{j,\pm,2}(E^{\prime})dE^{\prime}.
\end{equation}
For given $i,\pm$ and $j,\pm$ this integral is finite and continuous in $E$ when $E$
stays away from $E_{i,\pm}+E_{j,\pm}$, so let $E=E_{i,\pm}+E_{j,\pm}+\epsilon$ with
$\epsilon$ small. Now we make the variable transformation
$E^{\prime}=E_{j,\pm}+\epsilon^{\prime}$. Then \eqref{int2} may be written as
\begin{equation}
\label{int3}
\int\Theta(\mp(\epsilon-\epsilon^{\prime}))\Theta(\mp\epsilon^{\prime})
|\epsilon-\epsilon^{\prime}|^{-1/2}|\epsilon^{\prime}|^{-1/2}
g_1(\epsilon-\epsilon^{\prime})g_2(\epsilon^{\prime})d\epsilon^{\prime},
\end{equation}
where $g_1(\epsilon-\epsilon^{\prime})=f_{i,\pm,2}(\epsilon-\epsilon^{\prime}
+E_{i,\pm})$ and $g_2(\epsilon^{\prime})=f_{j,\pm,2}(\epsilon^{\prime}+E_{j,\pm})$
are smooth and vanish outside a neighborhood of the origin. Also the choice of the
sign in $\mp$ in the first Heaviside step function refers to $i,\pm$ while the choice
of the sign in $\mp$ in the second refers to $j,\pm$. For $\epsilon\neq 0$
\eqref{int3} is continuous in $\epsilon$. Also both limits $\epsilon\downarrow 0$ and
$\epsilon\uparrow 0$ exist but are in general different. In fact, consider the case
+,+ of the signs, i.e., the choice $i,-$ and $j,-$. Then \eqref{int3} vanishes for
$\epsilon <0$ and for $\epsilon
>0$ we make the variable transformation $\epsilon^{\prime}=\epsilon x$ such
that \eqref{int3} takes the form
\begin{equation*}
\int_{0}^{1}(1-x)^{-1/2}x^{-1/2}
g_1(\epsilon(1-x))g_2(\epsilon x)dx,
\end{equation*}
which is continuous in $\epsilon >0$ and has a finite limit equal to
\begin{equation*}
g_1(0)g_2(0)\int_{0}^{1}(1-x)^{-1/2}x^{-1/2}dx
\end{equation*}
when $\epsilon\downarrow 0$. The other three cases are discussed similarly. This
concludes the proof of our claim on the differentiability of $\enn(E;H^{(n=2)})$. To
see that $\enn(E;H^{(n=3)})$ has a derivative which is H\"{o}lder continuous of index
$1/2$, we use the representation \eqref{conv4} with $n=3$ and $l=1$ for the
derivative of $\enn(E;H^{(n=3)})$. Since we just established that
$\enn^{\prime}(E;H^{(n=2)})$ is continuous in $E$ except for a discrete set of
discontinuities (without finite accumulation points) some easy arguments establish
this last claim.

\end{appendix}


\end{document}